\documentclass[%
 reprint,
 superscriptaddress,
 amsmath,amssymb,
 aps,
 pre,
nofootinbib,
]{revtex4-1}

\usepackage[usenames,dvipsnames]{xcolor}
\usepackage{amsthm}
\usepackage{amssymb}

\usepackage{diagbox}

\usepackage{graphicx}% Include figure files
\usepackage{dcolumn}% Align table columns on decimal point
\usepackage{bm}% bold math
\usepackage[unicode, colorlinks, urlcolor=blue, linkcolor=blue, pagecolor=blue, citecolor=blue]{hyperref} 
\usepackage[T2A]{fontenc}
\usepackage[utf8]{inputenc}
\usepackage[russian,english]{babel}
\usepackage{comment}
\addto\captionsenglish{}
\usepackage{bigstrut}
\usepackage{float}

\let\oldcfrac\cfrac
\renewcommand{\cfrac}[2]{\oldcfrac{#1}{#2}\,}
\usepackage{comment}
\usepackage{hhline}
\begin{document}

\title{One--Component Plasma of a Million Particles via angular--averaged Ewald potential: A Monte Carlo study}

\author{G. S. Demyanov}
\affiliation{Joint Institute for High Temperatures, Izhorskaya 13 Bldg 2, Moscow 125412, Russia}
\affiliation{Moscow Institute of Physics and Technology, Institutskiy Pereulok 9, Dolgoprudny, Moscow Region, 141701, Russia}
\author{P. R. Levashov}
\affiliation{Joint Institute for High Temperatures, Izhorskaya 13 Bldg 2, Moscow 125412, Russia}
\affiliation{Moscow Institute of Physics and Technology, Institutskiy Pereulok 9, Dolgoprudny, Moscow Region, 141701, Russia}

\date{\today}

\begin{abstract}
In this work, we derive a correct expression for the one--component plasma (OCP) energy via
the angular--averaged Ewald potential (AAEP).
Unlike E.~Yakub and C.~Ronchi (J. Low Temp. Phys. 139, 633 (2005)), who had tried to obtain the same energy expression from a two--component plasma model, we used the original Ewald potential for an OCP. A constant in the AAEP was determined using the cluster expansion in the limit of weak coupling. 
The potential has a simple form suitable for effective numerical simulations. To demonstrate the advantages of the AAEP, we performed a number of Monte--Carlo simulations for an OCP with up to a million particles in a wide range of the coupling parameter. Our computations turned out at least two orders of magnitude more effective than those with a traditional Ewald potential. A unified approach is offered for the determination of the thermodynamic limit in the whole investigated range. Our results are in good agreement with both theoretical data for a weakly coupled OCP and previous numerical simulations. We hope that the AAEP will be useful in path integral Monte Carlo simulations of the uniform electron gas.
\end{abstract}

\maketitle

\section{Introduction}
A one--component plasma (OCP) is an important and well--studied model with a long history \cite{Hansen:1973,Slattery:1980,Baus:PR:1980,Stringfellow:1990,Potekhin:2000,Khrapak:2014,Caillol:2010}. An OCP is usually defined as a system of point ions immersed in a uniform neutralizing background \cite{Brush:1966}. An OCP is a good approximation of a two--component plasma (TCP) if the ions can be considered classical (their de Broglie wavelength is much less than the interionic distance) and the electrons are highly degenerated. Such conditions correspond to the interior of Jupiter (for hydrogen ions), white dwarfs (for helium, carbon and oxygen ions), neutron star crust (e.g., for iron ions) and can also be obtained in laser experiments \cite{Baus:PR:1980}. A quantum--mechanical analogue of the OCP is the well--known ``jellium'' model introduced by Wigner in 1938 \cite{Wigner:TFS:1938} and received much attention in the last decade \cite{Dornheim:PR:2018, Filinov:PRE:2015, Dornheim:PRE:2019, Dornheim:JCP:2019}. 

Contrary to a \emph{classical} TCP, an OCP satisfies the so-called $H$-stability condition classically; the thermodynamic limit exists for its thermodynamic properties both in the microcanonical and canonical ensembles \cite{Lieb:JSP:1975}. Another advantage of the OCP is the dependence of all its properties only on a dimensionless coupling parameter $\Gamma$
\cite[Sec. II]{Hansen:1973}, \cite[\textsection 31, Problem 1]{Landau:StatPhys}, \cite{Brush:1966}. 

On the other hand, the long--range Coulomb interaction between the ions should be taken into account. This causes significant difficulties both in analytical and numerical studies. In particular, the electrostatic energy of an infinite OCP is a conditionally convergent series so the summation result depends on the order of the terms \cite{UnderstandingMolecularSimulation}. For crystalline Coulomb systems, the solution to this problem was proposed by Ewald \cite{Ewald:1921}. Adding to and subtracting from the original sum a system of normally--distributed screening charges, one may transform the sum into two rapidly converging series. This procedure defines an anisotropic short--ranged Ewald potential \cite{Brush:1966, Hansen:1973, Baus:PR:1980}, which is used in further analysis. 

In computer simulations, a cubic box with periodic boundary conditions is often used. So a system without a long--range order acquires a translational symmetry. The Ewald's summation technique is also valid in this case; therefore, almost all Monte Carlo (MC) or molecular dynamics (MD) studies of an OCP are based upon this approach. 

Many different approaches are developed to implement the Ewald's technique in computations. The traditional Ewald summation \eqref{eq:spherEwald}, \eqref{eq:angularEwald} requires the computational cost of the order $O(N^2)$. The  grid--based Ewald methods, such as particle--mesh Ewald \cite{Darden:1993}, smoothed particle--mesh Ewald \cite{Essmann:1995} and particle--particle--particle mesh \cite[Chapter 8]{Hockney2021}, allow to reduce the complexity to $O(N\log N)$. In our work and most of the ones cited below, the traditional Ewald approach is used.

In 1966, the first extensive numerical investigation of an OCP was carried out by Brush, Sahlin and Teller using a MC method \cite{BST:1966}.
The authors calculated thermodynamic properties and
radial distribution functions (RDFs) in the range $0.05~\leq~\Gamma~\leq~100$, which corresponds to the fluid state.  However, a small number of particles $N~\leq~500$ was used, being especially critical at $\Gamma \ll 1$ when the Debye length becomes large \cite{Caillol:2010}; the thermodynamic limit was not considered. 

In 1973, the famous work by Hansen \cite{Hansen:1973} was published, in which the equation of state of a fluid OCP was constructed using MC simulation data in the range $1\leq\Gamma\leq160$. The number of particles in the cell was also small, $N\leq 250$; the thermodynamic limit was not investigated. The anisotropic part of the Ewald potential was approximated by an optimized expansion in Kubic harmonics; its accuracy was criticized in Ref.~\cite[Sec. II]{Slattery:1980}. Also, a detailed examination shows that there are some misprints in the formulas in Ref.~\cite{Hansen:1973} (e.g., Eq. (7) or (B3)).

In Ref.~\cite{Slattery:1980}, Slattery, Doolen and DeWitt used a more precise approximation for the Ewald potential compared to \cite{Hansen:1973, BST:1966}. The authors examine both the fluid and solid phases of an OCP ($1~\leq\Gamma\leq~300$). Using $N = 128$, they built the equation of state for both phases and estimated the fluid--solid transition point as $\Gamma_m = 168\pm 4$. The $N$-dependence was considered by the same authors only in the following paper \cite{Slattery:1982}. In Ref.~\cite{Slattery:1982}, a linear dependence on $N^{-1}$ is assumed. The thermodynamic limit was found by the extrapolation of MC data. Nevertheless, the authors state that they do not know the correct dependence on $N$, although the proposed dependence fits the MC data well. In Ref.~\cite{Slattery:1982}, the authors present  a significantly different value of $\Gamma_m = 178\pm 1$. 

As we see, the convergence of the results on $N$ is the major question  of all such studies, including the calculation of $\Gamma_m$. This important aspect is discussed in \cite{Ogata:1987}, where $N\leq 1024$ was used. The ``center-of-mass correction'' of the OCP energy is considered. This correction was applied, e.g., in \cite[Eq. (16)]{Hansen:1973}, \cite{Slattery:1982}. The authors conclude that this correction cannot be justified for the fluid OCP but is necessary for the solid phase.

Significant progress in the accuracy of MC simulation results was made in the works by J. M. Caillol \emph{et al.} \cite{caillol:1981,Caillol:1982,Caillol:Compar:1999, Caillol:1999,Caillol:2010, J_M_Caillol_2020}. Starting from \cite{caillol:1981}, they developed a method for modeling an OCP on the sphere surface of different dimensions. They demonstrated in detail the ability of this approach by simulating a 2D OCP \cite{Caillol:1982} using $N\leq 256$ in the $0.5~\leq\Gamma\leq~200$ region. In \cite{Caillol:Compar:1999}, Caillol discusses the difference between simulations in a cubic cell and on a sphere surface from a theoretical point of view.

Using this method, the most reliable result for the OCP energy was obtained in \cite{Caillol:1999} for $1~\leq\Gamma\leq~190$ and in \cite{Caillol:2010} for $0.1~\leq\Gamma\leq~1$. The authors consider the thermodynamic limit in both papers; $N\leq3200$ is used in \cite{Caillol:1999} and $N~\leq~51200$ in \cite{Caillol:2010}. To reduce the statistical error, a record number of configurations ($10^8$--$10^9$) were used in MC simulations.

It can be seen from the above that over time, more and more attention is paid to the problem of the $N$-dependence in OCP MC simulations. Modern computers made it possible to increase the number of particles in OCP MC simulations from $N=10^2$ in 1966 \cite{BST:1966} to $N=5\times 10^4$ in 2010 \cite{Caillol:2010}, respectively, as well as significantly decrease the statistical error. We also see that the statement in \cite[bottom of p. 19]{Baus:PR:1980} about a weak dependence of the OCP energy on $N$ is simply wrong.

As for the MD, it is mainly applied to investigate the dynamic and transport properties of an OCP \cite{Scheiner:2019, J_M_Caillol_2020, Korolov:2015}, including in the presence of external fields \cite{Ott:2011, Dubey:2018, Bonitz:2022}. The first reference to such a simulation was given in \cite{BST:1966}, and the first ``extensive'' study was done by Hansen \emph{et al.} \cite{Hansen:1974, Hansen:1975}. In these works, the authors present the computations of the velocity autocorrelation function and the dynamical structure factor. The thermal conductivity and viscosity of an OCP are calculated in \cite{Bernu:1978} using the Kubo formula. The OCP internal energy can also be obtained from MD simulations \cite{Farouki:1993}, as well as the estimations of the fluid--solid transition point $\Gamma_m$ \cite{Ueshima:1997}.

Using MD simulations one may investigate the OCP system with the Yukawa potential (YOCP) \cite{Salin:2003, Ott:2014}, including  the low screening limit \cite{Farouki:9885:1994}. In such a regime the YOCP results tend to the OCP ones \cite{Hamaguchi:9876:1994}. 

Two main theoretical approaches for an OCP have been developing: the theory of integral equations for an RDF, $g(r)$, and the expansion of the internal energy on $\Gamma$. Starting from the Debye and H\"{u}ckel \cite{Debye:1923} result, in which the expansion of the internal energy up to the term $\Gamma^{3/2}$ was obtained, there has been significant progress. In Ref.~\cite{Cohen:1969}, the diagram technique was used for this purpose. According to Caillol \cite{Caillol:2010}, further development of Ortner \cite{Ortner:1999} produced no significant difference from the result of \cite{Cohen:1969}. Nevertheless, these studies have made the expansion applicable at $\Gamma~\leq~0.3$, whereas the result by Debye and H\"{u}ckel is reliable only at $\Gamma\leq 0.01$. The work by Brilliantov \cite{Brilliantov:1998} derives the approximate ``first-principle'' OCP equation of state, demonstrating the error of 2-5\% in the range $0\leq \Gamma\leq 250$. 

Nevertheless, the expansion on $\Gamma$ does not produce relatively accurate results even at $\Gamma = 1$ \cite[Fig. 1]{Caillol:2010}, \cite[p. 27]{Baus:PR:1980}. At higher $\Gamma$, the theory of integral equations for an RDF was applied. In particular, the hypernetted chain (HNC) approximation \cite{Ng:1974} gives quite accurate results for the energy at $\Gamma \le 1$, as it was shown in \cite[Fig. 1]{Caillol:2010} and \cite[Sec. 3.7]{Baus:PR:1980}. 
As well as for an OCP, this theory can be implemented for a YOCP, using the Rogers--Young, Ballone--Pastore--Galli--Gazzillo and variational modified HNC \cite{Castello:CPP:2021} and the soft mean spherical \cite{Tolias:PRE:2014} approximations. 
However, at stronger interaction, integral equations should be corrected to agree with MC or MD results \cite{Rogers:1983,Iyetomi:1992,Castello:2022}.

Thus, despite the development of various theoretical methods, numerical modeling is indispensable in the study of an OCP. As a consequence, there is a request for fast and efficient calculation methods that can produce highly accurate results. To reduce the amount of computations in the usual Ewald technique, E.~Yakub and C.~Ronchi \cite{Yakub:2003} proposed an angular--averaged Ewald potential (AAEP) for the simulation of an isotropic Coulomb TCP. Indeed, as the Ewald potential is anisotropic, it seems redundant to apply it to isotropic plasma systems. The AAEP was correctly presented for a TCP without a clear derivation \cite{Yakub:2003}. It was successfully used in computations \cite{Yakub:JPA:2006,Jha:2010,Filinov:PRE:2020, Yakub:2007, Fukuda:2011, Fukuda:2012, Guerrero:2011, Fukuda:2013, Guo:2011, Lytle:2016, Nikitin:2020, KAMIYA201326}, including the simulation of quantum systems \cite{Filinov:PRE:2015, Dornheim:PRE:2019, Dornheim:JCP:2019}. Recently we have succeeded to find a step by step derivation of the AAEP for a TCP \cite{Demyanov:arxiv:2022}.  
E.~Yakub and C.~Ronchi computed the OCP energy from the TCP one; the AAEP in case of an OCP was not presented. However, the energy expression given in \cite[Eq. (8)]{Yakub:2005}, \cite[Eq. (8)]{Yakub:JPA:2006} is wrong and has no step by step derivation either. This confusion stimulated us to derive a correct expression for the OCP AAEP by the direct averaging of the Ewald potential \eqref{eq:Ewaldfull}. This paper is devoted to solving this problem. Also we obtain the correct expression for the OCP energy. Finally, we perform MC simulations for a fluid OCP using up to a \emph{million} particles to obtain precise values of energy.

The paper is organized as follows. Section~\ref{sec:av} contains the problem statement and the derivation of the AAEP for the case of the OCP. In Section~\ref{sec:corr} we make the correction of the AAEP and obtain the correct expression for the OCP energy. Section~\ref{sec:app} is devoted to the applications of the new potential including MC simulations of the OCP energy with up to a million particles in a wide range of the coupling parameter. The conclusion is formulated in Section~\ref{sec:conc}.

\section{\label{sec:av}Averaging the OCP Ewald potential}
In this section, we derive an expression for the AAEP in an OCP, examine its main properties and make it suitable for numerical calculations.

\subsection{Problem Statement}
Consider a cubic cell with a side length $L$. This cell contains $N$ positively charged particles at positions $\textbf{r}_i$, $i=1,\dots, N$. All the particles have equal charge $Ze > 0$, $e > 0$, and $Z$ is a charge number. Negative charge with density $-NZe/L^3$ uniformly distributed throughout the cell ensures electroneutrality. We call this cell the main cell. The cell repeats itself in the three mutually perpendicular directions; each $i$-th particle has an infinite number of images with positions $\textbf{r}_i+\textbf{n}L$. Here, $\textbf{n}$ is an integer vector $\textbf{n}=(n_x, n_y, n_z)$, $n_x, n_y, n_z \in \mathbb{Z}$. Then the negatively charged uniform background occupies the entire space with the same charge density. We call such an infinite electroneutral system the OCP. Now at some point $\textbf{r}\in \mathbb{R}^3$, we have the following charge density $w(\textbf{r})$:
\begin{equation}
w(\textbf{r}) = Ze\sum\limits_{i=1}^N\sum_{\textbf{n}}\delta(\textbf{r}-\textbf{r}_i-\textbf{n}L)-N\cfrac{Ze}{L^3}.
\end{equation}
Here, $\delta(\textbf{r}-\textbf{r}_i)$ is the Dirac $\delta$-function, $\sum_{\textbf{n}}$ denotes a summation over all integer vectors $\textbf{n}$.
An interaction potential between positively charged particles $\phi(\textbf{r})$ satisfies the Poisson equation (in Gaussian units):
\begin{equation}
\Delta \phi(\textbf{r}) = -4\pi w(\textbf{r}),
\end{equation}
where $\Delta$ is the Laplacian. 

The excess thermodynamic properties of the OCP depend only on the dimensionless parameter $\Gamma$ \cite[Sec. II]{Hansen:1973}, \cite[\textsection 31, Problem 1]{Landau:StatPhys}:
\begin{equation}
\label{eq:gammadef}
\Gamma = \cfrac{(Ze)^2}{k_BTa},
\end{equation}
where $a^3 \equiv \frac{3}{4\pi \rho}$ is the ``ion-sphere radius'', $\rho = N/L^3$ is the \emph{number} density, $T$ is the temperature of the system and $k_B$ is the Boltzmann constant. The value of $\Gamma$ denotes the ratio of a characteristic potential energy to a characteristic kinetic energy of particles. The $\Gamma \ll 1$ regime corresponds to a weakly--coupled OCP in which the kinetic energy dominates over potential one. From the definitions of $a$ and $\rho$, there is a useful relationship between $L$ and $N$: 
\begin{equation}
\label{eq:LandNrelation}
(L/a)^3 = 4\pi N/3.
\end{equation} 

We use $a$ as the unit of length in all subsequent formulas. 

The energy of the OCP can be written in terms of the effective interaction potential between particles in the main cell. This pair potential $v(\textbf{r})$ is given in \cite[Eq. (5)]{Hansen:1973}, \cite[Eqs. (11)-(13)]{Brush:1966}; it takes into account the presence of the negatively charged background. Then the ratio of the OCP total potential energy $E^{\text{Ew}}$ to $k_BT$, $U \equiv E/(k_BT)$, reads:
\begin{equation}
\label{eq:energy}
U^{\text{Ew}}(\Gamma) = U_0(\Gamma) + \cfrac{\Gamma}{2}\sum\limits_{i=1}^N\sum\limits_{\substack{j=1\\ j\neq i}}^Nv(\textbf{r}_{ij}),
\end{equation}
\begin{equation}
\label{eq:Ewaldfull}
v(\textbf{r}) = v_1(r) + v_2(\textbf{r}),
\end{equation}
\begin{equation}
\label{eq:spherEwald}
Lv_1(r) = \cfrac{\mathrm{erfc}(\sqrt{\pi}r/L)}{r/L} - 1,
\end{equation}
\begin{multline}
\label{eq:angularEwald}
Lv_2(\textbf{r}) = \sum_{\textbf{n}\neq\textbf{0}}\biggl[\cfrac{\mathrm{erfc}(\sqrt{\pi}|\textbf{r}/L+\textbf{n}|)}{|\textbf{r}/L + 
\textbf{n}|} \\
{} + \cfrac{e^{-\pi n^2}}{\pi n^2}\cos\left(\cfrac{2\pi}{L}\textbf{r}\cdot \textbf{n}\right)\biggr],
\end{multline}
where $\textbf{r}_{ij} = \textbf{r}_i - \textbf{r}_j$, $n = |\textbf{n}| = (n_x^2+n_y^2+n_z^2)^{1/2}$, $r = |\textbf{r}|$. The potential $v(\textbf{r})$ is called the Ewald potential. Summation $\sum_{\textbf{n}\neq\textbf{0}}$ means that the term $\textbf{n}=(0,0,0)\equiv \textbf{0}$ is omitted. The potential $v(\textbf{r})$ consists of a spherically symmetrical $v_1(r)$ and angular dependent $v_2(\textbf{r})$ parts. Since the right--hand sides of Eqs. \eqref{eq:spherEwald}, \eqref{eq:angularEwald} are dependent  on the ratio $\textbf{r}/L$ we introduce the notation:
\begin{equation}
\textbf{x}\equiv \textbf{r}/L, \quad x \equiv r/L.
\end{equation}

The energy expression \eqref{eq:energy} contains the summation over $N$ particles only. The interaction between all the periodic particle images is included in the Ewald potential \eqref{eq:angularEwald}. Thus, a particle in the main cell interacts only with $N-1$ other particles in the main cell or with the nearest ``images'' in one of the neighboring cells. Such a procedure corresponds to the ``minimum--image convention''. As a result, for fixed particle positions, each particle interacts with $N -1$ particles that are located in a cube centered at that particle \cite[Sec. III]{Brush:1966}, \cite[Sec. IV]{Demyanov:arxiv:2022}.

The first term $U_0$ in \eqref{eq:energy} denotes the self-interaction energy between particles with their replicas. This term \emph{depends on the potential} $v(\textbf{r})$ \cite[Eq. (10)]{Brush:1966}:
\begin{equation}
	\label{eq:e0pot}
U_0(\Gamma) = \cfrac{\Gamma}{2}N\lim\limits_{|\textbf{r}|\to 0}\left(v(\textbf{r}) - \cfrac{1}{r}\right) =  -\cfrac{1}{2}N^{2/3}\Gamma M_{sc}.
\end{equation}
Here, $M_{sc} = 1.760118884$ is the Madelung constant of a simple cubic lattice. The last equality in Eq. \eqref{eq:e0pot} is valid only for the Ewald potential \eqref{eq:Ewaldfull}. Note that the formula for the self-interacting term given in \cite{Hansen:1973} (see \cite[Eq. (7)]{Hansen:1973}) is false.

\subsection{Averaging the Ewald potential}
In disordered media, the angular dependence of $v_2(\textbf{r})$ leads to needless calculations since all orientations are equivalent. At first, we directly apply the approach of E. Yakub and C. Ronchi, which was used earlier for a TCP \cite{Yakub:2003, Demyanov:arxiv:2022}, to average $v_2(\textbf{r}_{ij})$ over all directions at a distance $r_{ij}$:
\begin{equation}
v_2^a(r_{ij}) = \cfrac{1}{4\pi}\int\limits_{-1}^1d(\cos \theta)\int\limits_0^{2\pi} v_2(\textbf{r}_{ij})d\psi.
\end{equation}
The module in the first term of $v_2(\textbf{r})$ in Eq. \eqref{eq:angularEwald} has the following form ($\textbf{n}\cdot \textbf{x} = nx\cos\theta$):
\begin{equation}
|\textbf{x}+\textbf{n}| = \sqrt{x^2+n^2+2nx\cos\theta}.
\end{equation}
Now we can integrate it over angles:
\begin{equation}
\cfrac{1}{2}\int\limits_{-1}^1d(\cos \theta)\cfrac{\mathrm{erfc}(\sqrt{\pi}|\textbf{x}+\textbf{n}|)}{|\textbf{x}+\textbf{n}|}\\=
\frac{f(|n-x|)-f(|n+x|)}{2\pi n x} ,
\end{equation}
where
\begin{equation}
\label{eq:fDef}
f(n) = e^{-\pi n^2} -\pi n\mathrm{erfc}\left(\sqrt{\pi } n \right).
\end{equation}
Averaging the second term in Eq. \eqref{eq:angularEwald}
\begin{equation}
\cfrac{1}{2}\int\limits_{-1}^1d(\cos \theta)\cos\left(2\pi x n \cos\theta\right) = \cfrac{\sin\left(2\pi x n \right)}{2\pi x n},
\end{equation}
we obtain the angular-averaged pair potential $v^a_2(x)$:
\begin{multline}
\label{eq:angularPotAveraged}
Lv^a_2(x) = \cfrac{1}{2\pi n x} \sum_{\textbf{n}\neq\textbf{0}}\left[
f(|n-x|)-f(|n+x|) 
\phantom{\cfrac{e^{-\pi n^2}}{\pi n^2}}\right.
\\\left.+ \cfrac{e^{-\pi n^2}}{\pi n^2}\sin(2\pi n x)
\right].
\end{multline}

Further we are going to consider only the case of $x~<~1$ (see the explanation below and \cite[Sec. IV]{Demyanov:arxiv:2022}). Since the minimum value of $n$ is $1$, we will reveal the module as follows: $\left| n-x\right| = n - x$. Next, we expand $v^a(x) = v_1(x) + v_2^a(x)$ into the converging series by $x$ at $x = 0$:
\begin{multline}
\label{eq:taylorSeries}
Lv^a(x) = \cfrac{\mathrm{erfc}(\sqrt{\pi}x)}{x} - 1 + Lv^a_2(x) 
\\=
\frac{1}{x}-C_0+\frac{2 \pi  x^2}{3}+\sum_{k=2}^{\infty}C_kx^{2k}.
\end{multline}
The coefficient $C_0$ is related to the Madelung constant of a simple cubic lattice, $M_{sc}$:
\begin{multline}
C_0 = 3-\sum_{\textbf{n}\neq\textbf{0}}\left(\frac{\mathrm{erfc}\left(\sqrt{\pi } n\right)}{n}+\frac{e^{  -\pi n^2}}{\pi 
	n^2}\right) 
\\=-\lim\limits_{\textbf{x}\to \textbf{0}} \left(Lv_2(\textbf{x})-\cfrac{1}{x}\right) = M_{sc}(4\pi/3)^{1/3}.
\end{multline}
The general formula for the series coefficients (for $k~\geq~1$) is derived in Appendix~\ref{app:coefs} and given in Eq. \eqref{eq:CoefsExpression}. We rigorously prove in Appendix~\ref{app:summation} that $C_k = 0$ for $k\geq 2$. Thus, the expression for the OCP AAEP is:
\begin{equation}
\label{eq:averagedPotM}
v^a(r) =\frac{1}{r}\left[1-M_{sc}\cfrac{r}{r_m} + \cfrac{1}{2}\left(\cfrac{r}{r_m}\right)^3\right],
\end{equation}
where
\begin{equation}
\label{eq:rmdefin}
r_m = \left(\cfrac{3}{4\pi}\right)^{\frac{1}{3}}L = N^{1/3} < L
\end{equation}
is the radius of the sphere $4\pi r_m^3/3=L^3$ with equivalent volume $L^3$. The last relation is obtained using Eq. \eqref{eq:LandNrelation}. Note that $L/2 < r_m$. The expression for AAEP in case of an OCP was not presented in the original works \cite{Yakub:2005, Yakub:JPA:2006}.  

The AAEP reaches its minimum value $v^a(r_m) = \left[\frac{3}{2}-M_{sc}\right]/r_m$ at $r = r_m$, $\left.\partial v^a(r)/ \partial r\right|_{r=r_m}  = 0$. For $L>r>r_m$, the potential increase; this behavior is incorrect (see \cite[Sec. IV]{Demyanov:arxiv:2022}). Thus, one must consider Eq. \eqref{eq:averagedPotM} up to a point $r = r_m$; we redefine $v^a(r)$ by zero for $r> r_m$:
\begin{equation}
\label{eq:zeroingPot}
v^a(r>r_m) = 0.
\end{equation} 
Each particle is affected by $N_s-1$ particles in the sphere of a radius $r_m$. Here, $N_s$ is the full number of particles in the sphere with a center at some ion; this number is dependent on the position (more details are given in our previous paper \cite[Secs. IV, V]{Demyanov:arxiv:2022}). Below we show that the average value of $\bar{N}_s$ during the MC simulation matches the value $N$ (see Fig. \ref{fig:interactionnumber}).

The term $-M_{sc}r/r_m$ in Eq. \eqref{eq:averagedPotM} is responsible for the background negative charge. Our calculations show that the coefficient $M_{sc}$ leads to the divergence of the potential energy per ion as the number $N$ increases. 
The reason for such behavior is assumed to be related to the angular averaging.  
Therefore, the coefficient $M_{sc}$ must be changed. Below we obtain the correct value for the coefficient from the cluster expansion. 

\section{\label{sec:corr}Correction and shifting of the potential}
In this section, we correct the constant coefficient in the AAEP and obtain the final OCP energy expression \eqref{eq:energyAv}, which can be used in applications.
\subsection{Correction of the AAEP}
Now we consider the AAEP \eqref{eq:averagedPotM} with an unknown coefficient: $M_{sc}\to C$.
The excess part of the canonical partition function for a system of $N$ particles interacting through potential \eqref{eq:averagedPotM} is \cite[Sec. III]{Hansen:1973}:
\begin{multline}
Q_N(\Gamma) = \int\dots\int\exp\left\{-\Gamma\sum_{i< j}v^a(r_{ij}) \right\}d^3r_1\dots d^3r_N 
\\= \int\dots\int\prod_{{i< j}}\exp\left\{-\Gamma v^a(r_{ij}) \right\}d^3r_1\dots d^3r_N.
\end{multline}
Here the integration region over each coordinate is a volume $L^3 = 4\pi N/3$. 
We define the Mayer $f$-function:
\begin{equation}
\exp\left\{-\Gamma v^a(r_{ij}) \right\} = 1 + f_{ij},
\end{equation}
and rewrite the partition function:
\begin{equation}
Q_N(\Gamma) = \int\dots\int\prod_{{i< j}}(1 + f_{ij})d^3r_1\dots d^3r_N.
\end{equation}
Providing the cluster expansion
\begin{equation}
\prod_{{i< j}}(1 + f_{ij}) = 1 + \sum_{i< j} f_{ij} + \dots,
\end{equation}
we get in the first order of $f_{ij}$:
\begin{multline}
Q_N(\Gamma) \approx \int\dots\int\left(1 + \sum_{i< j} f_{ij}\right)d^3r_1\dots d^3r_N 
\\= \left(\cfrac{4\pi}{3}N\right)^N + \sum_{i< j}\int\dots\int f_{ij}d^3r_1\dots d^3r_N.
\end{multline}
After the integration, the last term has the form:
\begin{multline}
\sum_{i< j}\int\dots\int f_{ij}d^3r_1\dots d^3r_N = 
-\cfrac{N(N-1)}{2}\left(\cfrac{4\pi}{3}N\right)^N 
\\+ \cfrac{N(N-1)}{2}\left(\cfrac{4\pi}{3}N\right)^{N-2}I(\Gamma, N),
\end{multline}
where  $I(\Gamma, N)$ is the following integral:
\begin{multline}
I(\Gamma, N)  =
\int d \textbf{r}_i\int\exp\left\{-\Gamma v^a(|\textbf{r}_j - \textbf{r}_i|) \right\} d(\textbf{r}_j - \textbf{r}_i)
\\=
\int d \textbf{r}_i\int\limits_0^{r_m}\exp\left\{-\Gamma v^a(u) \right\} 4\pi u^2 du 
\\= 
3 \left(\cfrac{4\pi}{3}N\right)^2 K(\Gamma, N),
\end{multline}
\begin{equation}
K(\Gamma, N) = \int\limits_0^1 \exp\left\{-\Gamma v^a(sr_m) \right\}s^2 ds,
\end{equation}
\begin{equation}
v^a(s r_m) = 
\cfrac{1}{r_m}\left(\cfrac{1}{s}- C + \cfrac{s^2}{2}\right).
\end{equation}
Thus, we get the final expression for the excess part of the canonical partition function in the first order of the cluster expansion:
\begin{equation}
Q_N(\Gamma) \approx \left(\cfrac{4\pi}{3}N\right)^N\left\{1+\cfrac{N(N-1)}{2}(3K(\Gamma, N) - 1)\right\}.
\end{equation}
The potential energy $E_{\text{cl}}$ then reads:
\begin{multline}
\cfrac{E_{\text{cl}}}{Nk_BT}(\Gamma) =
-\cfrac{1}{\Gamma}\cfrac{1}{Q_N(\Gamma)}\cfrac{\partial Q_N(\Gamma)}{\partial \Gamma} \\
\approx
-\cfrac{1}{\Gamma}\cfrac{1}{\cfrac{2/3}{N(N-1)}+K(\Gamma, N) - 1/3}\cfrac{\partial K(\Gamma, N)}{\partial \Gamma}.
\end{multline}
With $\Gamma \to 0$, the derivative
\begin{multline}
\cfrac{\partial K(\Gamma, N)}{\partial \Gamma} = 
- \cfrac{1}{r_m}\int\limits_0^1\left(\cfrac{1}{s}- C + \cfrac{s^2}{2}\right)
\\\times  \exp\biggl\{2\ln s  - \cfrac{\Gamma}{r_m}\left(\cfrac{1}{s}-C + \cfrac{s^2}{2}\right) \biggr\}
 ds
\end{multline}
must tend to zero since $\Gamma \to 0$ corresponds to the ideal gas and $E_{\text{cl}}/(Nk_BT)(0) = 0$ for any $N$. From this condition, we find the unknown value of $C$:
\begin{equation}
\label{eq:95const}
\cfrac{\partial K(\Gamma \to 0, N)}{\partial \Gamma} = 0 \Rightarrow C = 9/5.
\end{equation}
The final expression for the AAEP \eqref{eq:averagedPotM} should be used with $M_{sc}\to C = 9/5$.
The self--interaction energy $U_0$ also should be changed to $U_{0a}$ in accordance with \eqref{eq:e0pot}:
\begin{equation}
\label{eq:trueE0av}
U_{0a}(\Gamma) = \cfrac{1}{2}N\lim\limits_{r\to 0}\left(v^a(r) - \cfrac{1}{r}\right) =  -\cfrac{1}{2}N^{2/3}\Gamma C.
\end{equation}

\subsection{Shifting the potential}
Discontinuity of $v^a(r)$ at the point $r=r_m$ results in a discontinuity in the energy. Such behavior leads to different problems during numerical simulations \cite[p. 302]{Julian:2003}. Thus, we shift the average potential to make it zero at $r\geq r_m$. 
The contribution of the two-particle interaction to the internal energy is the following:
\begin{multline}
\label{eq:twoPartPotEn}
\cfrac{\Gamma}{2}\sum\limits_{i=1}^N\sum\limits_{\substack{j=1\\ j\neq i}}^N\left(v^a(r_{ij}) - v^a(r_m)+v^a(r_m)  \right) 
\\= \cfrac{N(N-1)\Gamma}{2}v^a(r_m) + 
\cfrac{\Gamma}{2}\sum\limits_{i=1}^N\sum\limits_{\substack{j=1\\ j\neq i}}^N\tilde{v}(r_{ij}),
\end{multline}
where the pair potential $\left. \tilde{v}(r)\right|_{r~\leq~r_m} = \left. v^a(r)\right|_{r~\leq~r_m} - v^a(r_m)$ has the form:
\begin{equation}
\label{eq:shiftedPot}
\tilde{v}(r) = 
\begin{cases}
\cfrac{1}{r}\left\{1 + \cfrac{1}{2}\left(\cfrac{r}{r_m}\right)\left[\left(\cfrac{r}{r_m}\right)^2-3\right]\right\}, &r < r_m\\
0, &r \geq r_m.
\end{cases}
\end{equation}

Since the number of ions in the sphere differs from $N$, we replace $\sum_{j=1\atop j\neq i}^{N}\tilde{v}(r_{ij})$ to $\sum_{j=1\atop j\neq i}^{N_{s,i}}\tilde{v}(r_{ij})$ in \eqref{eq:twoPartPotEn}. Here, $N_{s,i}$ is the number of ions in the sphere centered at an $i$-th ion.
The full potential energy then reads:
\begin{equation}
\label{eq:energyAv}
U(\Gamma) = \tilde{U}_0(\Gamma) +
\cfrac{\Gamma}{2}\sum\limits_{i=1}^N\sum\limits_{\substack{j=1\\ j\neq i}}^{N_{s,i}}\tilde{v}(r_{ij}).
\end{equation}
The constant term in \eqref{eq:energyAv} 
\begin{multline}
\tilde{U}_0(\Gamma) = U_{0a}(\Gamma) + \cfrac{N(N-1)\Gamma}{2}v^a(r_m)
=
 -\cfrac{N^{2/3}\Gamma C}{2} 
\\
+ \cfrac{N(N-1)}{2 r_m}\left[\frac{3}{2}-C\right]\Gamma
 = -\cfrac{3}{20}N^{2/3}\Gamma(N+5)
\end{multline}
includes both the effects of self-interaction and shifting of AAEP. Here, we used $C = 9/5$ \eqref{eq:95const} and $r_m = N^{1/3}$ \eqref{eq:rmdefin}. 

So, to calculate the OCP energy, one needs to evaluate the interaction of any ion placed in the center of a sphere with a radius $r_m$ with all other ions inside this sphere  and then sum up all such interactions. Each $i$-th particle  interacts only with $N_{s,i} - 1$ particles that are located in a sphere centered at the $i$-th particle.

One may notice that the radius of interaction $r_m~>~L/2$; thus, the influence of some ions should be taken twice. This was explained in more detail in \cite[Sec. III]{Yakub:2003}, \cite[Sec. 2]{Jha:2010}, \cite[Sec. V]{Demyanov:arxiv:2022}.

The potential \eqref{eq:shiftedPot} can also be applied in MD simulations. Since $\tilde{v}(r)$ is a smooth function, we calculate a force $\textbf{f}(\textbf{r})$ between two particles as follows:
\begin{multline}
\textbf{f}(\textbf{r}) = -\cfrac{(Ze)^2}{a^2} \nabla  \tilde{v}(r)  \\ =
\cfrac{(Ze)^2}{a^2}\textbf{r}\times
\begin{cases}
\cfrac{1}{r^3} - \cfrac{1}{r_m^3}, &r < r_m\\
0, &r \geq r_m,
\end{cases}
\end{multline}
where $\nabla$ is the gradient operator.

In Ref.~\cite[Eq. (8)]{Yakub:2005}, the formula for the OCP potential energy is given without derivation. For clarity, we reproduce the wrong formula \cite[Eq. (8)]{Yakub:2005} here (in the Gaussian units):
\begin{equation}
\label{eq:yakub}
U^{\text{(OCP)}}_{N} = - 0.9 \cfrac{NQ^2}{r_m} + \cfrac{1}{2} \sum_{i = 1}^N\sum_{\substack{j = 1\\ j\neq i}}^NQ^2\tilde{v}(r_{ij}).
\end{equation}
It has the same form as Eq. \eqref{eq:energyAv}, but the constant term $\tilde{U}_0(\Gamma)$ is incorrect. 
Nevertheless, E. Yakub and C. Ronchi published correct results for the Madelung constants and the OCP energy. 
The correct formula \eqref{eq:energyAv} with a rigorous derivation from the Ewald potential is the main result of this paper.

In Ref.~\cite{Demyanov:arxiv:2022} we derive the TCP energy expressed through the averaged TCP Ewald potential. Replacing in this formula the summation over negatively charged particles to the integration over a uniformly distributed background, it is possible to obtain the OCP energy \eqref{eq:energyAv}. We provide such a derivation in Appendix~\ref{app:derivocptcp}.

\section{\label{sec:app}Applications}
Below we demonstrate accurate calculations for the OCP with the AAEP. We calculate the Madelung constants for two lattices, examine the performance of the calculation algorithm, and present the results of MC simulations with \emph{one million} particles.

\subsection{Madelung constant}
In this subsection, we use the AAEP \eqref{eq:averagedPotM} to calculate the Madelung constant $M$ of body--centered cubic (bcc) and face--centered cubic (fcc) lattices to verify our procedure of energy calculation. The expression for $M$ in case of \eqref{eq:energyAv} takes the following form:
\begin{equation}
\label{eq:madelungAv}
M = -\cfrac{3}{20}N^{-1/3}(N+5)+
\cfrac{1}{2}\sum\limits_{\substack{j=1\\ j\neq i}}^{N_s}\tilde{v}(r_{ij}).
\end{equation}
The results can be seen in Tables~\ref{tab:bccMad}, \ref{tab:fccMad} for bcc and fcc lattices, respectively. Even though the number of ions in the sphere $N_s$ differs from $N$ and the ionic configuration is not spherically symmetric at all, we observe the convergence of $M$ with increasing $N$ for both lattices.   Note, that $N_s$ as well as $M$ is independent of $i$.

\begin{table}[h!]
	\centering
	\caption{Madelung constant for bcc-lattice calculated by Eq. \eqref{eq:madelungAv} with increasing number of ions $N$. $N_c$ denotes the number of primitive cells in the supercell, $N = 2N_c^3$.}
	\begin{tabular}{rrrll}
		\hline
		\multicolumn{1}{|c|}{$N_c$} & \multicolumn{1}{c|}{$N$} & \multicolumn{1}{c|}{$N_s-N$} & \multicolumn{1}{c|}{$M$} & \multicolumn{1}{c|}{Difference, \%} \bigstrut\\
		\hline
		\multicolumn{1}{|c|}{1} & \multicolumn{1}{c|}{2} & \multicolumn{1}{c|}{-1} & \multicolumn{1}{c|}{-0.8333856} & \multicolumn{1}{c|}{-6.98088} \bigstrut\\
		\hline
		\multicolumn{1}{|c|}{3} & \multicolumn{1}{c|}{54} & \multicolumn{1}{c|}{5} & \multicolumn{1}{c|}{-0.9036126} & \multicolumn{1}{c|}{0.85759} \bigstrut\\
		\hline
		\multicolumn{1}{|c|}{4} & \multicolumn{1}{c|}{128} & \multicolumn{1}{c|}{9} & \multicolumn{1}{c|}{-0.8998543} & \multicolumn{1}{c|}{0.43810} \bigstrut\\
		\hline
		\multicolumn{1}{|c|}{8} & \multicolumn{1}{c|}{1024} & \multicolumn{1}{c|}{-59} & \multicolumn{1}{c|}{-0.8941086} & \multicolumn{1}{c|}{-0.20322} \bigstrut\\
		\hline
		\multicolumn{1}{|c|}{17} & \multicolumn{1}{c|}{9826} & \multicolumn{1}{c|}{15} & \multicolumn{1}{c|}{-0.8956311} & \multicolumn{1}{c|}{-0.03327} \bigstrut\\
		\hline
		\multicolumn{1}{|c|}{37} & \multicolumn{1}{c|}{101306} & \multicolumn{1}{c|}{243} & \multicolumn{1}{c|}{-0.8959880} & \multicolumn{1}{c|}{0.00655} \bigstrut\\
		\hline
		\multicolumn{1}{|c|}{79} & \multicolumn{1}{c|}{986078} & \multicolumn{1}{c|}{-543} & \multicolumn{1}{c|}{-0.8959281} & \multicolumn{1}{c|}{-0.00013} \bigstrut\\
		\hline
		\multicolumn{1}{|c|}{171} & \multicolumn{1}{c|}{10000422} & \multicolumn{1}{c|}{203} & \multicolumn{1}{c|}{-0.8959254} & \multicolumn{1}{c|}{-0.00043} \bigstrut\\
		\hline
		\multicolumn{1}{|c|}{369} & \multicolumn{1}{c|}{100486818} & \multicolumn{1}{c|}{763} & \multicolumn{1}{c|}{-0.8959294} & \multicolumn{1}{c|}{0.00002} \bigstrut\\
		\hline
		\hline
		\multicolumn{3}{r}{Exact:} & \multicolumn{2}{l}{-0.8959293} \bigstrut[t]\\
	\end{tabular}%
	\label{tab:bccMad}%
\end{table}%

\begin{table}[h!]
	\centering
	\caption{Madelung constant for fcc-lattice calculated by Eq. \eqref{eq:madelungAv} with increasing number of ions $N$. $N_c$ denotes the number of primitive cells in the supercell, $N = 4N_c^3$.}
	\begin{tabular}{rrrll}
		\hline
		\multicolumn{1}{|c|}{$N_c$} & \multicolumn{1}{c|}{$N$} & \multicolumn{1}{c|}{$N_s-N$} & \multicolumn{1}{c|}{$M$} & \multicolumn{1}{c|}{Difference, \%} \bigstrut\\
		\hline
		\multicolumn{1}{|c|}{1} & \multicolumn{1}{c|}{4} & \multicolumn{1}{c|}{-3} & \multicolumn{1}{c|}{-0.8504467} & \multicolumn{1}{c|}{-5.07068} \bigstrut\\
		\hline
		\multicolumn{1}{|c|}{2} & \multicolumn{1}{c|}{32} & \multicolumn{1}{c|}{11} & \multicolumn{1}{c|}{-0.8971610} & \multicolumn{1}{c|}{0.14370} \bigstrut\\
		\hline
		\multicolumn{1}{|c|}{4} & \multicolumn{1}{c|}{256} & \multicolumn{1}{c|}{-7} & \multicolumn{1}{c|}{-0.8947975} & \multicolumn{1}{c|}{-0.12012} \bigstrut\\
		\hline
		\multicolumn{1}{|c|}{8} & \multicolumn{1}{c|}{2048} & \multicolumn{1}{c|}{45} & \multicolumn{1}{c|}{-0.8962085} & \multicolumn{1}{c|}{0.03738} \bigstrut\\
		\hline
		\multicolumn{1}{|c|}{17} & \multicolumn{1}{c|}{19652} & \multicolumn{1}{c|}{-175} & \multicolumn{1}{c|}{-0.8957744} & \multicolumn{1}{c|}{-0.01107} \bigstrut\\
		\hline
		\multicolumn{1}{|c|}{37} & \multicolumn{1}{c|}{202612} & \multicolumn{1}{c|}{89} & \multicolumn{1}{c|}{-0.8958525} & \multicolumn{1}{c|}{-0.00235} \bigstrut\\
		\hline
		\multicolumn{1}{|c|}{79} & \multicolumn{1}{c|}{1972156} & \multicolumn{1}{c|}{-169} & \multicolumn{1}{c|}{-0.8958673} & \multicolumn{1}{c|}{-0.00070} \bigstrut\\
		\hline
		\multicolumn{1}{|c|}{171} & \multicolumn{1}{c|}{20000844} & \multicolumn{1}{c|}{-207} & \multicolumn{1}{c|}{-0.8958733} & \multicolumn{1}{c|}{-0.00003} \bigstrut\\
		\hline
		\multicolumn{1}{|c|}{369} & \multicolumn{1}{c|}{200973636} & \multicolumn{1}{c|}{505} & \multicolumn{1}{c|}{-0.8958739} & \multicolumn{1}{c|}{0.00003} \bigstrut\\
		\hline
		\hline
		\multicolumn{3}{r}{Exact:} & \multicolumn{2}{l}{-0.8958736} \bigstrut[t]\\
	\end{tabular}%
	\label{tab:fccMad}%
\end{table}%
We observe the convergence and reach the accuracy of 6 significant digits for $\sim~10^9$ particles in both lattices. 
This analysis confirms the usability of the AAEP to calculate the energy of even \emph{ordered} structures. The accuracy may be improved even more with the increase of the single parameter---the number of ions $N$.

\subsection{Performance}
\begin{figure}[h!]
\centering
\includegraphics[width=1\linewidth]{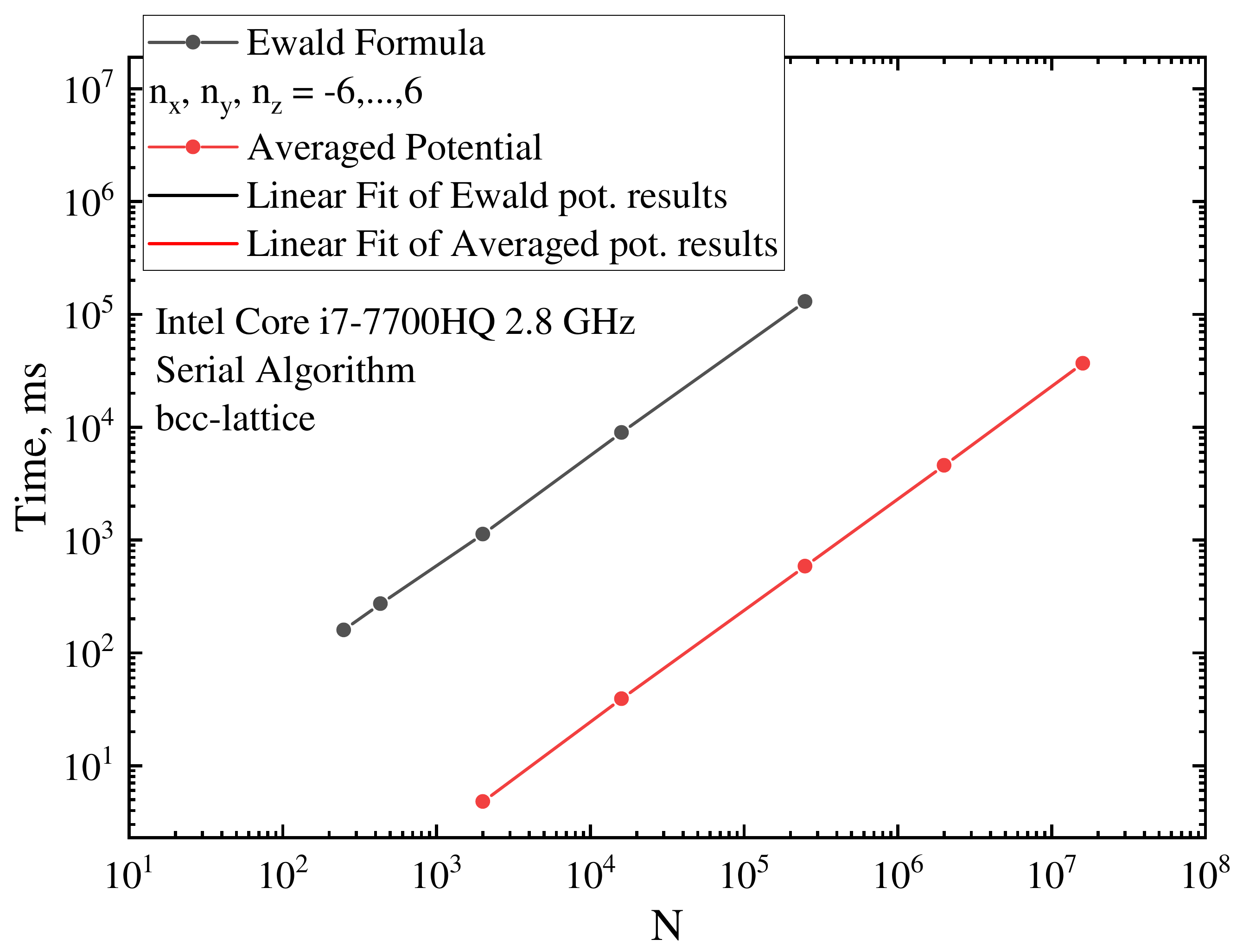}
\caption{The calculation time of the Madelung constant for a bcc-lattice as a function of $N$. Both methods are linear in $N$; the calculation
with the AAEP is 230 times faster than with the Ewald potential. The slope of the curve obtained with the Ewald
potential is $0.521$ ms, obtained with the averaged one is $2.39 \times 10^{-3}$ ms.}
\label{fig:performance}
\end{figure}

It is of practical importance to compare the performance of computations between the exact Ewald formula and the AAEP. 
To demonstrate the calculation performance via the AAEP, we calculate the Madelung constant of the bcc-lattice for a different number of particles $N$ in the supercell and measure the computational time. For comparison, we obtain $M$ using the Ewald \eqref{eq:Ewaldfull} and averaged \eqref{eq:shiftedPot} potentials. For the Ewald potential, we have taken into account several terms in the sum over $\mathbf{n}$ with $n_x$, $n_y$, $n_z = -6,\ldots ,6$. All the calculations were sequential and performed with a CPU Intel Core i7-7700HQ 2.8 GHz. The results are shown in Fig.~\ref{fig:performance}.
Both dependencies are linear as $N$ increases. The ratio of the slopes is $\approx 230$; it means that the calculation of $M$ is 230 times faster with the AAEP than with the Ewald potential. In addition, $v^a(r)$ is independent of any external parameters. This makes it  perspective for numerical simulations of Coulomb systems.

The calculation of the sum $\sum_{j=1\atop j\neq i}^{N_{s,i}}\tilde{v}(r_{ij})$ in Eq.~\eqref{eq:energyAv} can be parallelized. We simply distribute the terms of this sum over various processes during an MC simulation, that allows us to effectively speed up the computation. A similar approach was considered in \cite{Jha:2010} for TCP calculations.

Next, we perform MC simulations of the OCP in a wide range of parameter $\Gamma$. We demonstrate the possibility of a simulation with \emph{1 million particles} in the supercell.
\subsection{MC simulations}
We performed MC simulations using \eqref{eq:energyAv} for $\Gamma = 0.01, 0.05, 0.1, 1, 10, 100$. We demonstrate the capabilities of the AAEP using a \emph{million} particles in a simulation cell for $\Gamma = 0.01$--$1$. Such an amount of particles is important since the Debye length $\lambda_D~=~1/\sqrt{3\Gamma}$ diverges as $\Gamma\to 0$. Also, to obtain a reliable thermodynamic limit, it is necessary to know the value of energy at a significantly large $N$. Therefore one should consider a large simulation cell to obtain correct data. 

\subsubsection{Technique}
A very good description of a MC simulation procedure can be found in \cite[Sec. IV]{Brush:1966} and \cite[Chapter 5]{SadusBook:1999}. Here we briefly describe the algorithm that we used.

We start from an initial ion configuration of $N$ particles in a cubic supercell with random positions and calculate its energy $U_{init}$ by Eq. \eqref{eq:energyAv}. Then some ion $i$ at a position $\textbf{r}_i$ is randomly chosen. Next, we calculate the potential $u(\textbf{r}_i) = \sum_{j=1\atop j\neq i}^{N_{s,i}}\tilde{v}(r_{ij})$ at a point $\textbf{r}_i$ created by other particles (in the sphere of radius $r_m$). Now, this particle is moved as it is described in \cite{Brush:1966} producing a new ion position $\textbf{r}'_i$. Again, we calculate the potential energy $u(\textbf{r}'_i)$ and the total change in energy $\Delta U = \left[u(\textbf{r}'_i)-u(\textbf{r}_i)\right]\Gamma$. This trial move is accepted according to the Metropolis \emph{et al.} algorithm \cite{Metropolis:1953} as it described in \cite{Brush:1966}. Now we have a new ionic configuration with energy $U_{init} + \Delta U$; some ion is randomly selected again and the whole procedure repeats itself. Thus, we obtain a sequence $U(m)$, where $m$ denotes the ionic configuration number (or the trial move number). 
During the simulation we maintain the acceptance rate near 50\% (if it is possible) according to Algorithm 5.1 in \cite{SadusBook:1999}. It is stated in \cite[Sec. 5.1.2, p. 192]{SadusBook:1999} that such an algorithm increases the efficiency of a simulation.

Since we start from a randomly distributed ions position, which is not an equilibrium ionic configuration for a given $\Gamma$ and $N$, we need to discard the starting section of the simulation. After reaching the equilibrium, we perform $m_{tot} = 10^7$ steps for $N = 10^2, 10^3, 10^4, 10^5, 10^6$ at $\Gamma = 0.01, 0.05, 0.1, 1$. To decrease the statistical error, at $\Gamma = 10, 100$ we performed $m_{tot} = 10^8$ MC steps with $N = 10^2$--$10^5$. We also had to perform simulations for $N = 150$, $\Gamma = 100$ because the result for $N=100$ is out of normal dependence (see Fig. \ref{Fig:Ndependence}, d). As only large enough $N$ can be used in a low $\Gamma$ regime, we performed simulation for $N = 5\times 10^4$ and $\Gamma = 0.01$. All the simulation parameters are collected in Table \ref{tab:techparams}.
 \begin{table}[h!]
   \centering
   \caption{Parameters of MC simulations with AAEP}
     \begin{tabular}{|c|c|c|c|}
     \hline
     \multicolumn{1}{|c|}{$\Gamma$} & $N$ & $m_{tot}$ & \multicolumn{1}{c|}{$n_b$} \bigstrut\\
	 \hline
	 $0.01$ & $10^2$, $10^3$, $10^4$, $5\times 10^4$, $10^5$, $10^6$ & $10^7$ & $5$ \bigstrut\\
     \hline
	 $0.05$ & $10^2$, $10^3$, $10^4$, $10^5$, $10^6$ & $10^7$ & $5$ \bigstrut\\
     \hline
     $0.1$ & $10^2$, $10^3$, $10^4$, $10^5$, $10^6$ & $10^7$ & $5$ \bigstrut\\
     \hline
     $1$ & $10^2$, $10^3$, $10^4$, $10^5$, $10^6$ & $10^7$ & $5$ \bigstrut\\
     \hline
     $10$ & $10^2$, $10^3$, $10^4$, $10^5$ & $10^8$ & $50$ \bigstrut\\
     \hline
     $100$ & $10^2$, $150$, $10^3$, $10^4$, $10^5$ & $10^8$ & $50$ \bigstrut\\
     \hline
     \end{tabular}%
   \label{tab:techparams}%
 \end{table}%

\begin{figure}[h]
\centering
\includegraphics[width=1\linewidth]{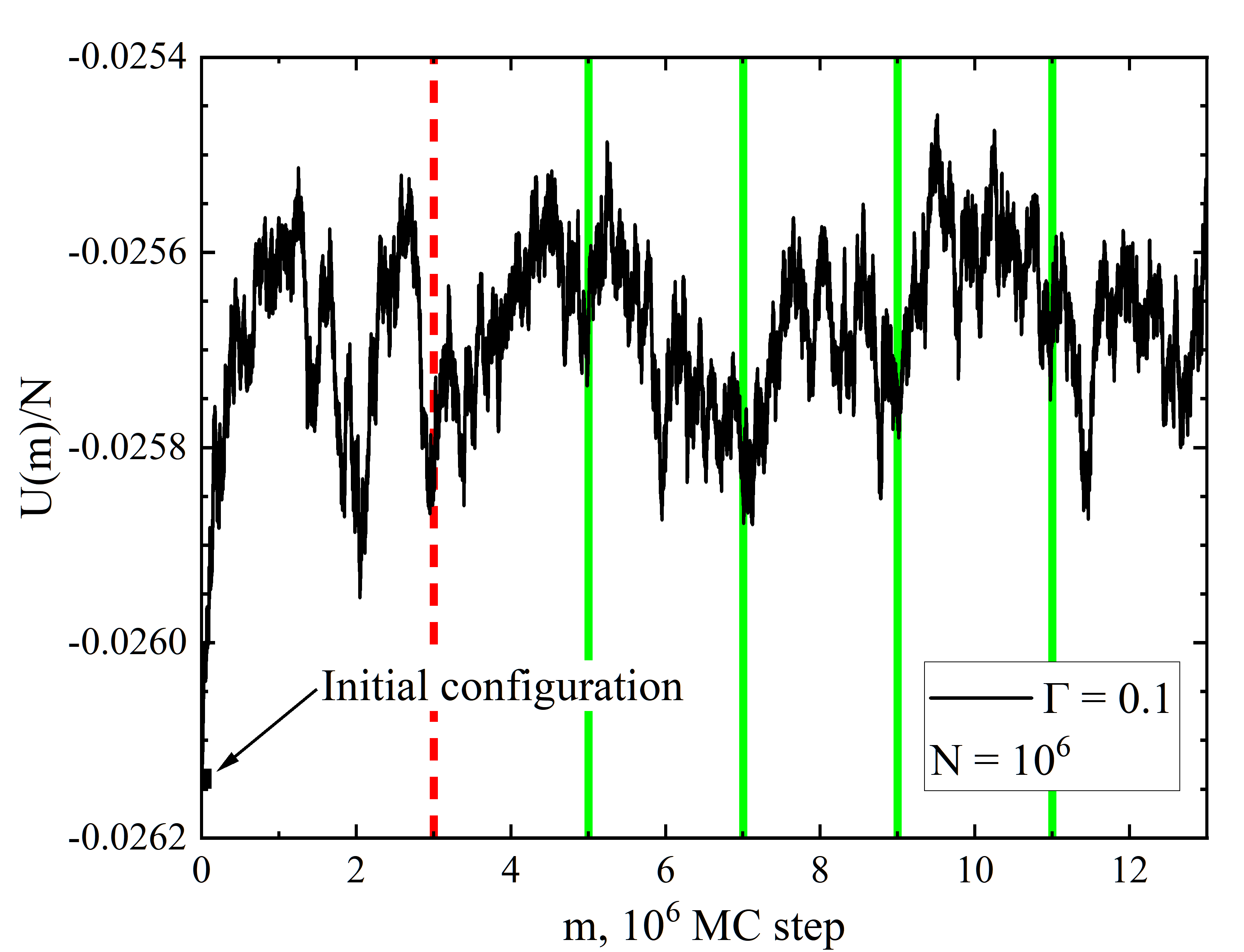}
\caption{The OCP energy dependence $U(m)/N$ with $10^6$ particles at $\Gamma = 0.1$. The vertical dashed red line shows the beginning of the  equilibrium section. Green solid vertical lines show the partitioning of the equilibrium section into blocks. 
}
\label{fig:energymln01}
\end{figure}

To calculate the averaged energy value and its statistical error, we use the standard block averaging \cite[Chapter 11.4]{GouldHarvey1996Aitc}. The entire equilibrium section is divided into several $n_b$ blocks (see Fig. \ref{fig:energymln01}). Each such block contains $m_{tot}/n_b = 2\times 10^6$ values of energy. We calculate an average energy value for each block, obtaining $n_b$ energy values $\bar{U}(l)$. The average energy value $U/N$ is calculated as the average of $\bar{U}(l)$.
The statistical error is estimated as the root of the variance of these mean values:
\begin{equation}
\label{eq:statError}
\sigma = \sqrt{\cfrac{1}{n_b - 1}\sum_{l=1}^{n_b}\left(\cfrac{\bar{U}(l)}{N} - \cfrac{U}{N}\right)^2}.
\end{equation}

To obtain the limit $N\to \infty$, which is often called the thermodynamic limit, we fit the dependence $U/N$ on $1/N$ and evaluate the limit as $1/N\to 0$. In the works \cite{Caillol:1999, Caillol:2010}, it is stated that with $N\to\infty$, the dependence $U/N (1/N)$ has the following form:
\begin{equation}
\cfrac{U}{N}(1/N) - \cfrac{U}{N}(0) \sim (1/N)^{2/3},\ \Gamma\to 0.
\end{equation}
For $\Gamma \gg 1$ the authors of \cite{Caillol:1999, Caillol:2010} use a linear fit $U/N(1/N) = U/N(0) + b\left(1/N\right)$ and a quadratic analog. In our work, we use the following fitting function \emph{for all} $\Gamma$ and $N$:
\begin{equation}
\label{eq:fitFunc}
\cfrac{U}{N}(1/N) = \cfrac{U}{N}(0) + b\left(1/N\right)^{\gamma}.
\end{equation}
The values $U/N(0), b, \gamma$ are obtained from the fitting. We do not provide any theoretical background for this approximation; nevertheless, it works well. The obtained values of $U/N(0)$ are shown in Table \ref{tab:addlabel}.

Also, we perform simulations with the traditional Ewald potential \eqref{eq:Ewaldfull} generating $m_{tot} = 10^7$ configurations for all $\Gamma = 0.1$--$100$ and $N = 10^2$--$10^4$ to compare.  Only the terms with $n_x, n_y, n_z = -6,\ldots,6$ are taken into account during the calculation of the sum over $\textbf{n}$ in Eq. \eqref{eq:angularEwald}. The results are presented in Table \ref{tab:addlabel1ewald} and in Figure \ref{Fig:Ndependence}.

It is not difficult to reach the equilibrium section with $N~\leq~10^5$. For one million particles, we used the following technique to speed up this simulation stage.  

To reach equilibrium with $N = 10^6$, we generated the initial configuration as follows. We performed an additional simulation for a small cell with $N = 15625$ to obtain one equilibrium configuration. Then we constructed a larger cell consisting of $2^3$ small cells, that resulted in an intermediate cell of $N = 125000$ ions. Repeating this procedure, we got an initial cell with $N = 10^6$ particles; it consisted of $2^6 = 64$ identical small cells with $N = 15625$. Nevertheless, the equilibration process required quite a long simulation. We provide a graph of $U(m)$ for $\Gamma = 0.1$ and $N = 10^6$ (Fig.~\ref{fig:energymln01}).
\noindent One can see that at least $\sim 10^6$ steps should be discarded. We discarded $3\times 10^6$ first steps.

\begin{figure}[h]
\centering
\includegraphics[width=1\linewidth]{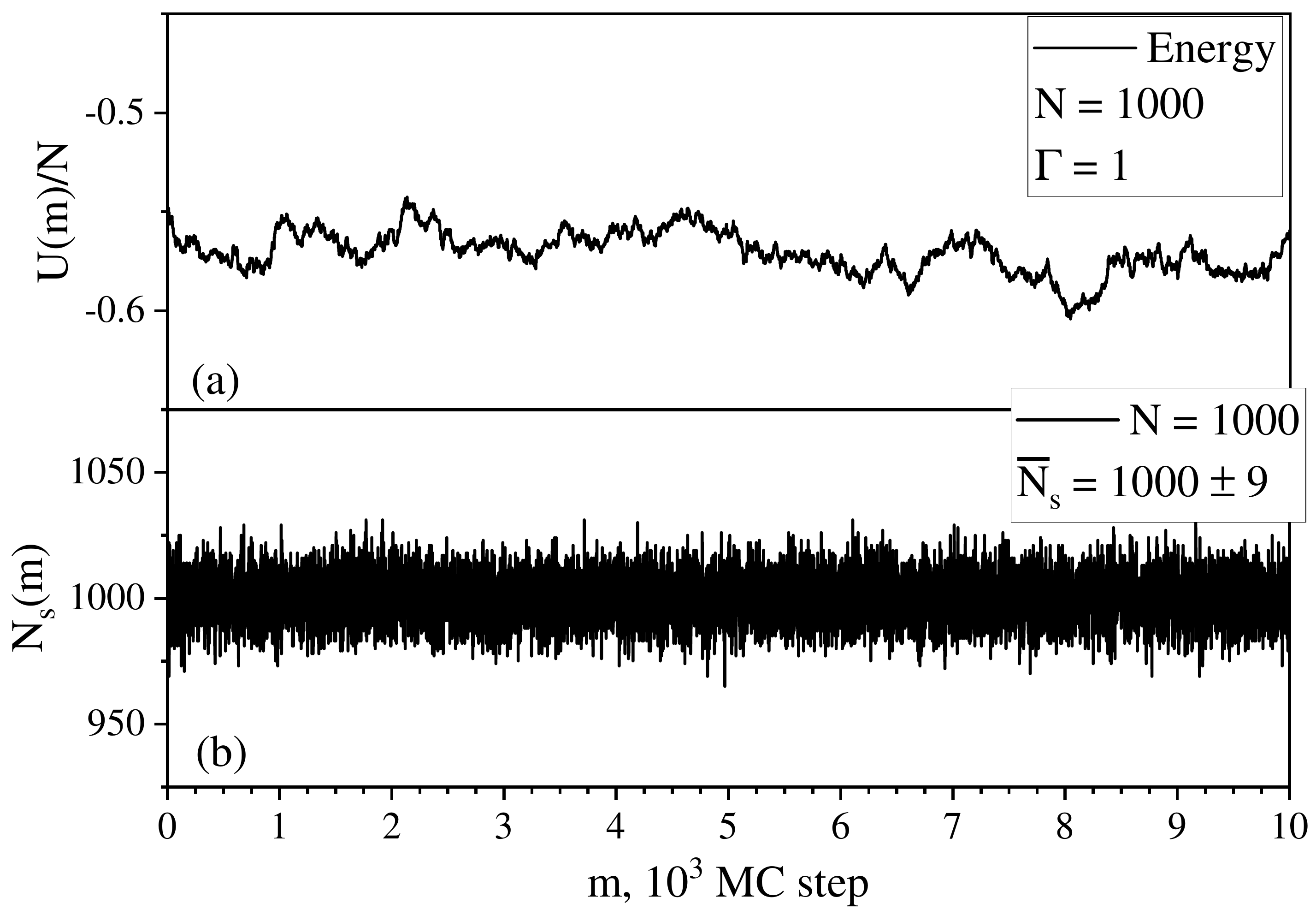}
\caption{The number of ions in a spherical cell and energy on the equilibrium section of MC simulation, $N = 10^3$, $\Gamma = 1$. (a) Energy. (b) Number of ions in a spherical cell.}
\label{fig:interactionnumber}
\end{figure}

\subsubsection{Results and discussions}
\begin{figure*}[ht!]
(a)~\includegraphics[width=0.46\linewidth]{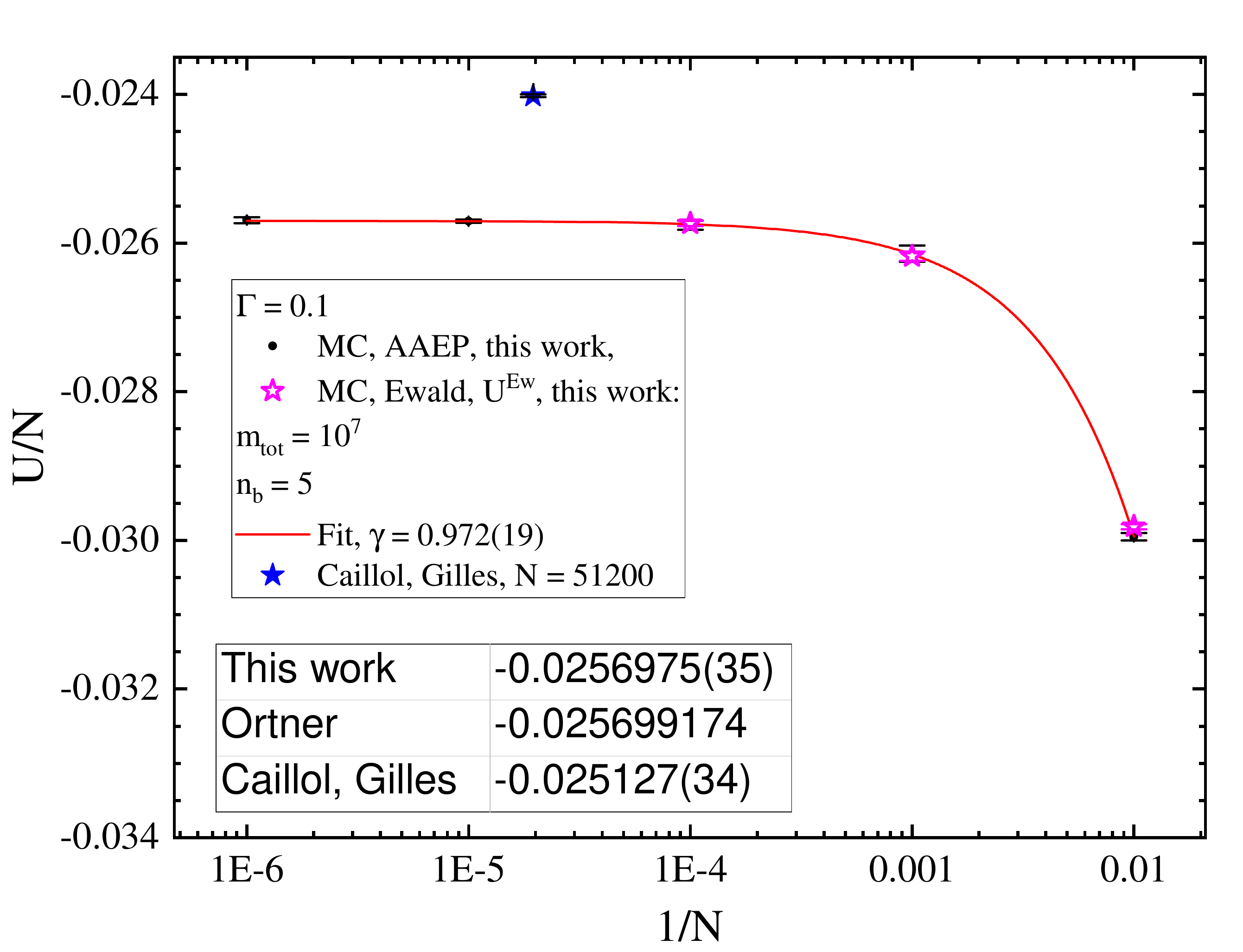}
(b)~\includegraphics[width=0.46\linewidth]{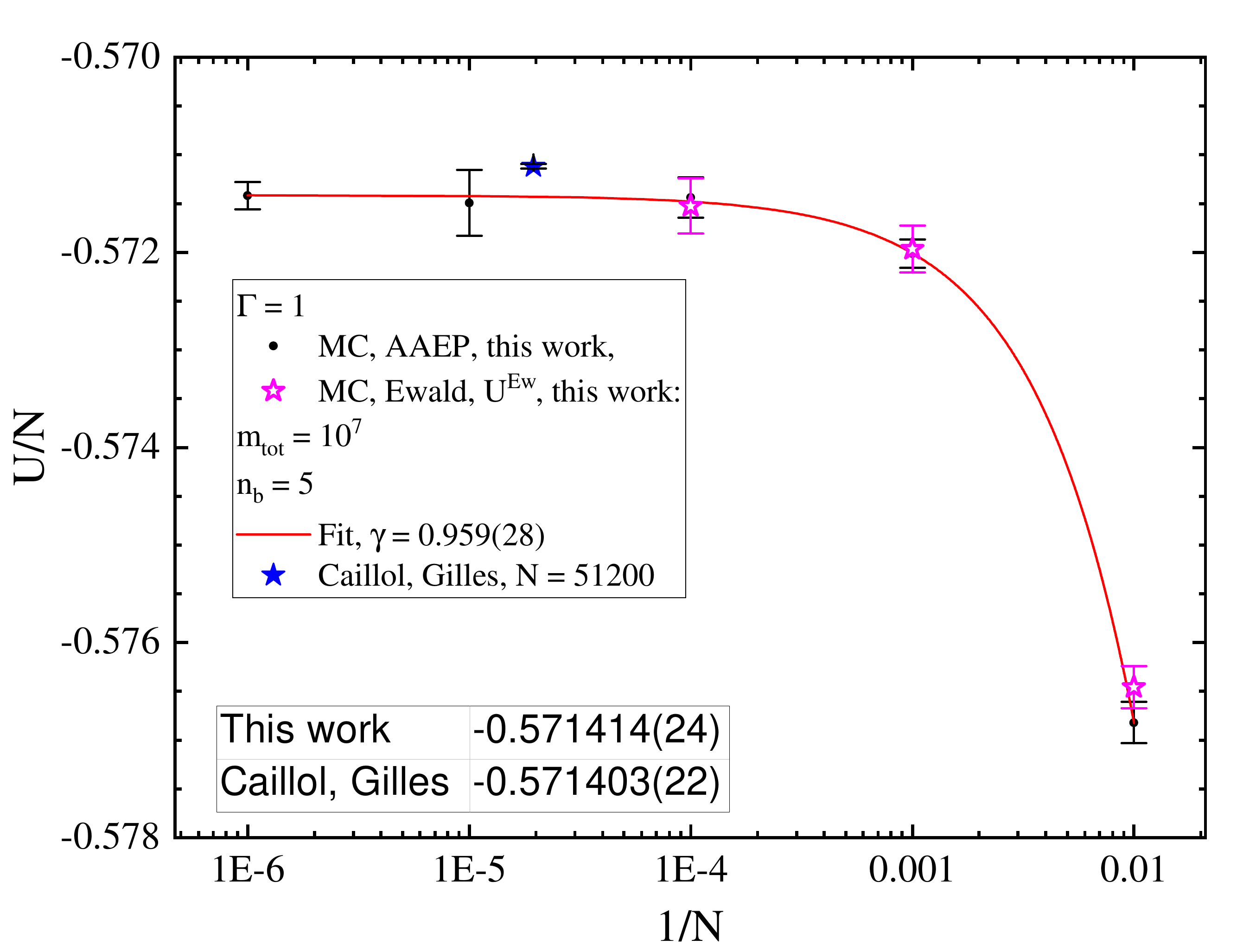}
(c)~\includegraphics[width=0.46\linewidth]{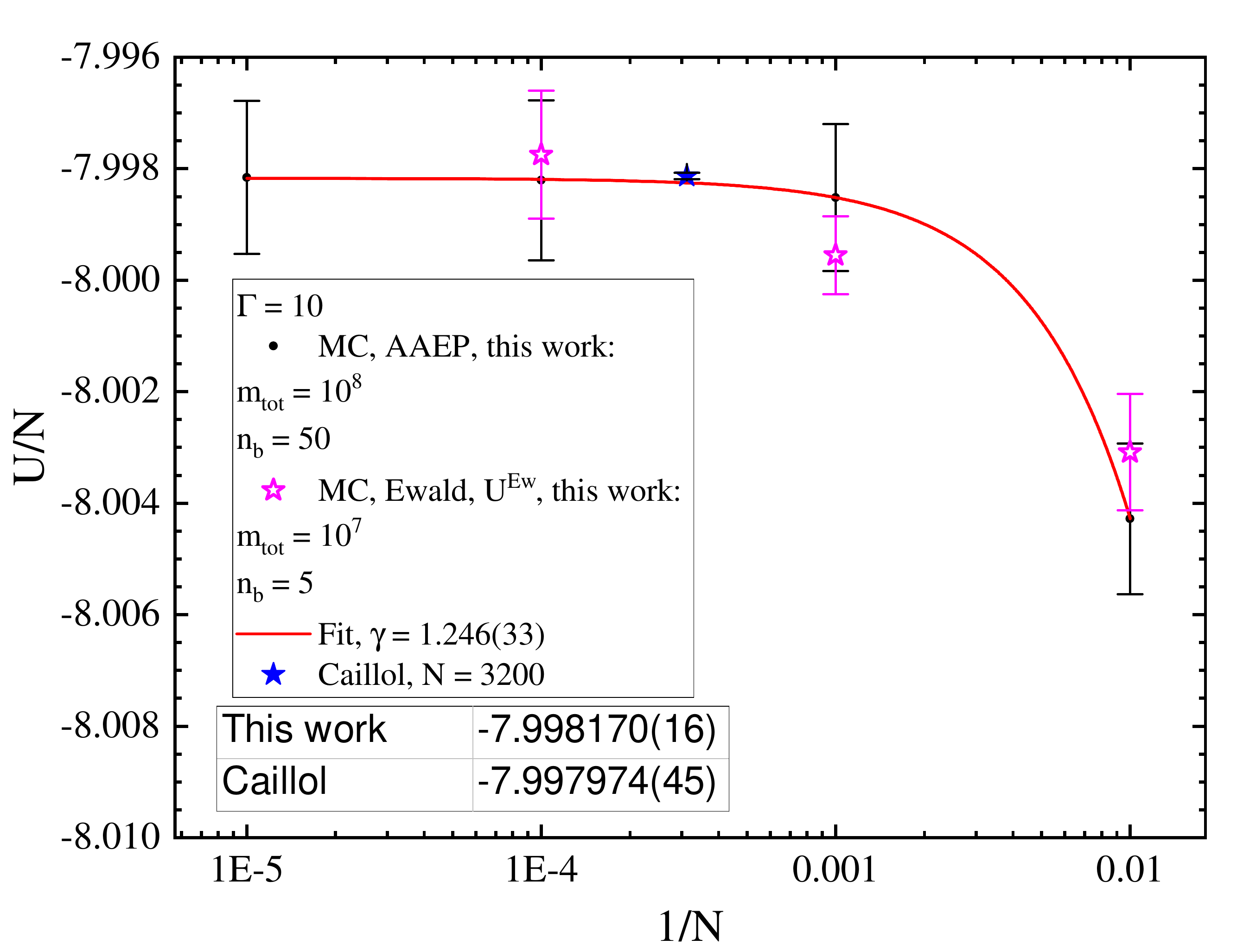}
(d)~\includegraphics[width=0.46\linewidth]{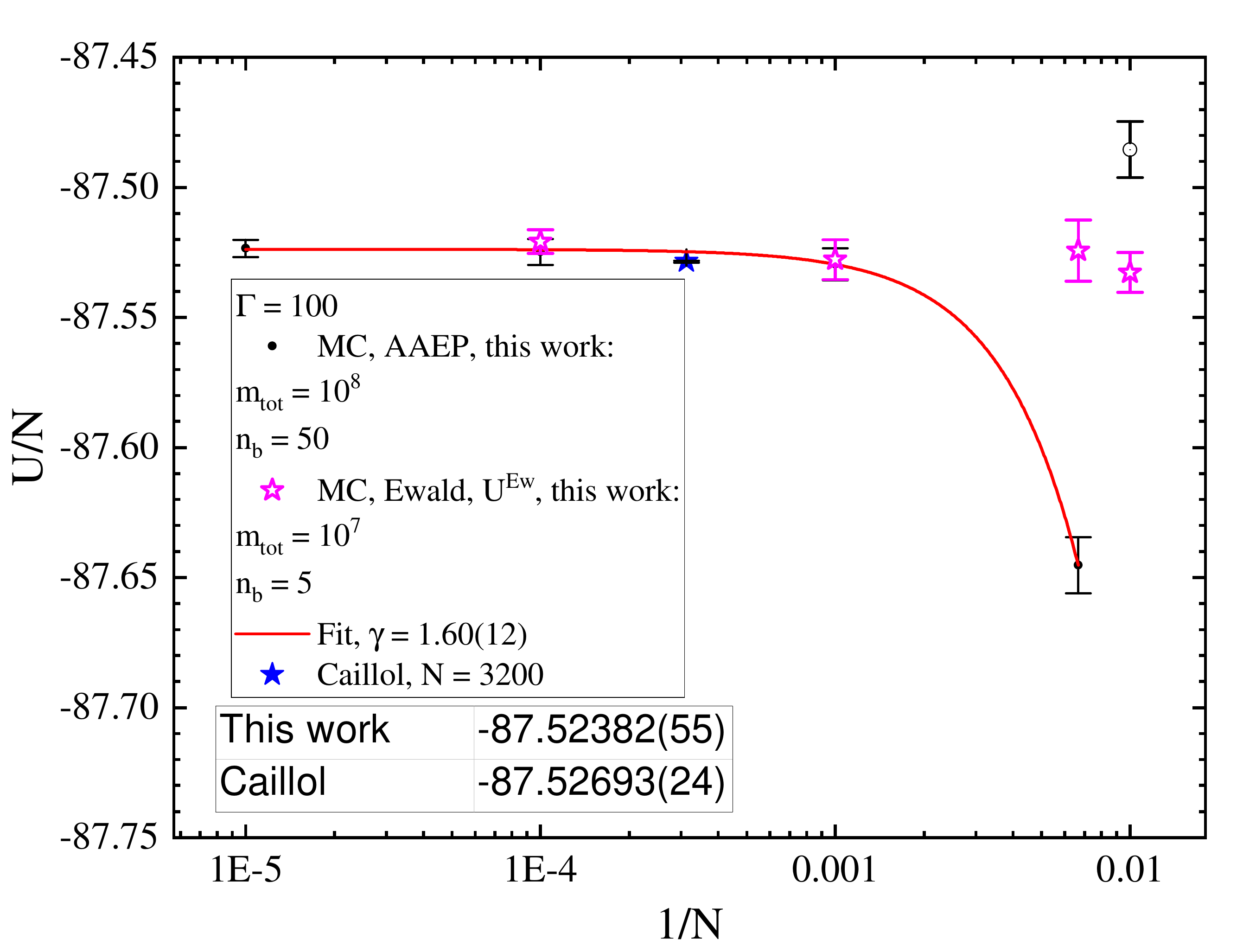}
\caption{The results of MC simulations at $\Gamma = 0.1, 1, 10, 100$ (points with bars) together with the numerical fit \eqref{eq:fitFunc} (solid lines). The error bars are calculated by Eq.~\eqref{eq:statError}. 
The thermodynamic limit energy values from \cite{Caillol:1999, Caillol:2010} are shown for comparison. The blue stars represent MC data for the largest $N$ in \cite{Caillol:1999, Caillol:2010}. The purple stars correspond to our MC simulations with the Ewald potential \eqref{eq:energy}. The digits in brackets correspond to one standard deviation.
}
\label{Fig:Ndependence}
\end{figure*}

First, we consider the number of interactions during MC simulation.

As we mentioned earlier, the number of ions $N_s$ in the sphere around a chosen ion can be different from $N$. Let $N_s(m)$ be the number of ions in a sphere with the center at a randomly chosen ion at its initial position, here $m$ is the trial move number. It turns out that the average number 
$\bar{N}_s =\frac{1}{m_{tot}}\sum_{m=1}^{m_{tot}}N_s(m)$ 
during MC simulations is close to $N$. The dependencies of $N_s(m)$ and $U(m)$ on the equilibrium section are shown in Fig.~\ref{fig:interactionnumber} for a simulation with $N = 10^3$, $\Gamma = 1$. In this simulation, $\bar{N}_s = (1.000~\pm~0.009)\times 10^3$ is close to $N = 10^3$. 

We obtained a numerical dependence of energy $U/N$ on the number of ions $N$ for $0.01\leq \Gamma \leq 100$ (see Tab. \ref{tab:addlabel1}). To calculate the thermodynamic limit, we use Eq.~\eqref{eq:fitFunc}. The results of our MC simulations and fitted curves are present in Table \ref{tab:addlabel1} and Figure \ref{Fig:Ndependence}. 

Paper \cite{Caillol:2010} gives the OCP energy values for $\Gamma = 0.1; 1$ obtained by solving the HNC integral equation; this method was outlined in \cite{Ng:1974}. According to \cite{Caillol:2010}, it is the most accurate theoretical result for the OCP energy in a low $\Gamma$ regime. In such a regime the Ortner $\Gamma$-expansion \cite{Ortner:1999} gives quite accurate results. For comparison, we provide MC results from \cite{Caillol:1999, Caillol:2010} and obtained by Ewald potential \eqref{eq:energy}.
The digits in brackets correspond to one standard deviation. 

Approximation \eqref{eq:fitFunc} quite accurately describes the dependence on $1/N$ over a wide range of $N$ and $\Gamma$, which vary by several orders of magnitude. However, for unknown reasons, at $\Gamma = 100$, the point $N = 100$ is out of dependence. This point was excluded from the fitting procedure.

At $\Gamma~\leq~1$, the power--law index changes weakly and is close to $\gamma = 1$. As $\Gamma$ increases, the index $\gamma$ also increases.

From approximation \eqref{eq:fitFunc}, we find energy values in the thermodynamic limit as $1/N \to 0$. Table \ref{tab:addlabel} presents the results of our extrapolation in comparison with the MC, HNC results by Caillol \emph{et al.} \cite{Caillol:1999, Caillol:2010} and Ortner expression \cite[Eq. (94)]{Ortner:1999}. 
For $\Gamma = 1$, our result is the same as that of Caillol \emph{et al.} In the case of $\Gamma = 0.01$--$0.1$, our result coincides with the HNC and Ortner values. It speaks in favor of our method in comparison to MC of \cite{Caillol:2010}. In the case of $\Gamma = 10$ and $\Gamma = 100$, we observe a difference of $2\times 10^{-3}\%$ and $4\times 10^{-3}\%$, respectively, from the Caillol result, which exceeds the statistical error. 

\begin{table*}[ht!]
\centering
\caption{MC results for $-U/N$ at $\Gamma = 0.1, 1, 10, 100$ as a function of $N$ using the AAEP. The digits in brackets correspond to one standard deviation. 
}
\label{tab:addlabel1}
\begin{tabular}{|c|ccccccc|} 
\hline
\diagbox{$\Gamma$}{$N$}                     & $10^2$       & $150$        & $10^3$        & $10^4$        & $5\times 10^4$   & $10^5$        & $10^6$         \\ 
\hline
$0.01$                                      & 0.0020724(45)& ---           & 0.0011814(96) & 0.000917(12) & 0.000875(11)  & 0.000866(15) & 0.000863(14)  \bigstrut[t]\\
$0.05$                                      & 0.012662(30) & ---           & 0.009855(38) & 0.009439(31) & ---  &  0.009405(54)&  0.009409(31)\bigstrut[t]\\
$0.1$                                      & 0.029951(50) & ---           & 0.026144(11) & 0.025756(63) & ---  & 0.025704(21) & 0.025691(37)  \bigstrut[t]\\
$1$                                        & 0.57682(21)  & ---           & 0.57201(15)  & 0.57144(21)  & ---  & 0.57149(34)  & 0.57142(14)  \bigstrut[t] \\
$10$                                       & 8.0043(14)   & ---           & 7.9985(13)   & 7.9982(14)   &  --- & 7.9982(14)   & ---            \bigstrut[t]  \\
$100$                                      & 87.485(11)   & 87.645(11) & 87.5295(62)  & 87.5248(50)  & ---  & 87.5235(33)  & ---             \bigstrut[b] \\
\hline
\end{tabular}
\end{table*}

\begin{table*}[ht!]
\centering
\caption{MC results for $-U^{\text{Ew}}/N$ at $\Gamma = 0.1, 1, 10, 100$ as a function of $N$ using the Ewald potential \eqref{eq:energy}, $n_x, n_y, n_z = -6,\ldots,6$. The digits in brackets correspond to one standard deviation. 
}
\label{tab:addlabel1ewald}
\begin{tabular}{|c|cccc|}
\hline
\diagbox{$\Gamma$}{$N$} & $10^2$       & $150$      & $10^3$       & $10^4$      \bigstrut[t] \\ \hline
$0.1$                   & 0.029815(34) & ---        & 0.026178(61) & 0.025737(35) \bigstrut[t]\\
$1$                     & 0.57646(21)  & ---        & 0.57197(24)  & 0.57153(28) \bigstrut[t] \\
$10$                    & 8.0031(11)   & ---        & 7.99955(70)  & 7.9977(12)  \bigstrut[t] \\
$100$                   & 87.5327(77)  & 87.524(12) & 87.5278(77)  & 87.5208(45) \bigstrut[t] \\ 
\hline
\end{tabular}
\end{table*}

\begin{table*}[ht!]
   \centering
   \caption{Thermodynamic limit of MC results for $-U/N$ at $\Gamma = 0.01, 0.05, 0.1, 1, 10, 100$. The results obtained using the AAEP agree very well with Caillol \emph{et al.} at $\Gamma = 1$, and with the theoretical values at $\Gamma = 0.01$--$0.1$. At $\Gamma = 10, 100$ we observe a small difference compared to \cite{Caillol:1999}. Third row represents the thermodynamic limit of data from \cite{Caillol:1999} obtained by Eq. \eqref{eq:fitFunc}. The digits in brackets correspond to one standard deviation.
}
     \begin{tabular}{|c|llllll|}
     \hline
     $\Gamma$& \multicolumn{1}{c}{0.01} & \multicolumn{1}{c}{0.05} & \multicolumn{1}{c}{0.1} & \multicolumn{1}{c}{1} & \multicolumn{1}{c}{10} & \multicolumn{1}{c|}{100} \bigstrut\\
     \hline
     MC (AAEP, this work) & 0.000 8611(42)\footnote{Only $N\geq 10^4$ were used for fitting  \eqref{eq:fitFunc}}  &0.009 395(13) & 0.025 6975(35) & 0.571 414(24) & 7.998 170(16) & 87.523 82(55) \bigstrut[t]\\
	 MC (Caillol \emph{et al.} \cite{Caillol:1999, Caillol:2010}) & \multicolumn{1}{c}{---}	 & \multicolumn{1}{c}{---}	& 0.025 127(34) & 0.571 403(22) & 7.997 974(45) & 87.526 93(24) \bigstrut[t]\\
	 Caillol's data, fit \eqref{eq:fitFunc}	 & \multicolumn{1}{c}{---}	 &\multicolumn{1}{c}{---}	 &	\multicolumn{1}{c}{---}			  & 0.571 654(88) & 7.997 996(39) & 87.524 42(75)\bigstrut[t] \\
	 Ortner \cite[Eq. (94)]{Ortner:1999} & 0.000 861 93 & 0.009 386 & 0.025 699 174 & 1.665 188  & \multicolumn{1}{c}{---} & \multicolumn{1}{c|}{---} \bigstrut[b]\\
	 Debye--H\"{u}ckel, $\sqrt{3}\Gamma^{3/2}/2$  & 0.000 866 03 & 0.009 682  & 0.027 386 128 & 0.866 025  & \multicolumn{1}{c}{---} & \multicolumn{1}{c|}{---} \bigstrut[b]\\
     HNC  & 0.000 861 98\footnote{Obtained from HNC fit \cite[Tab. 1]{Caillol:2010} \label{note1} } & 0.009 387$^{\text{\ref{note1}}}$ & 0.025 688 548 & 0.570 45534 & \multicolumn{1}{c}{---}  & \multicolumn{1}{c|}{---} \bigstrut[b]\\
     \hline
     \end{tabular}%
   \label{tab:addlabel}%
\end{table*}%

We fitted the Caillol \cite{Caillol:1999} MC data by Eq. \eqref{eq:fitFunc}; the result is presented in Table \ref{tab:addlabel} (third row). We observe, that for $\Gamma = 1, 100$ the energy value significantly changes. In contrast to the original Caillol result, these values coincide with ours at $\Gamma = 100$ and differ at $\Gamma = 1$. So the thermodynamic limit depends on the fitting function. Let us note that Caillol used different fitting functions at various $\Gamma$ (see Table II in \cite{Caillol:1999}). We think that the extrapolation $1/N\to 0$ with different dependencies on $N$ at various $\Gamma$ is a disadvantage of \cite{Caillol:1999}. In our paper, only one extrapolation dependence \eqref{eq:fitFunc} is used. 

In Figure \ref{Fig:Ndependence} we provide the MC data for a largest $N$ in \cite{Caillol:1999, Caillol:2010} (blue stars). Despite the fact that $\lambda^3_D$ is much less than the system volume, we believe that for the stated accuracy $N=3200$ is not enough. This circumstance may increase the final error in \cite{Caillol:1999}. 

The results of MC simulations via the Ewald potential is presented in Table \ref{tab:addlabel1ewald} and in Figure \ref{Fig:Ndependence} (purple stars). One can see that the traditional Ewald simulation and the one via the AAEP give close results for $N\geq 10^3$. At the same time, at $N=10^2$ there are differences, the biggest one is for $\Gamma = 100$.

\begin{figure}
\centering
\includegraphics[width=1\linewidth]{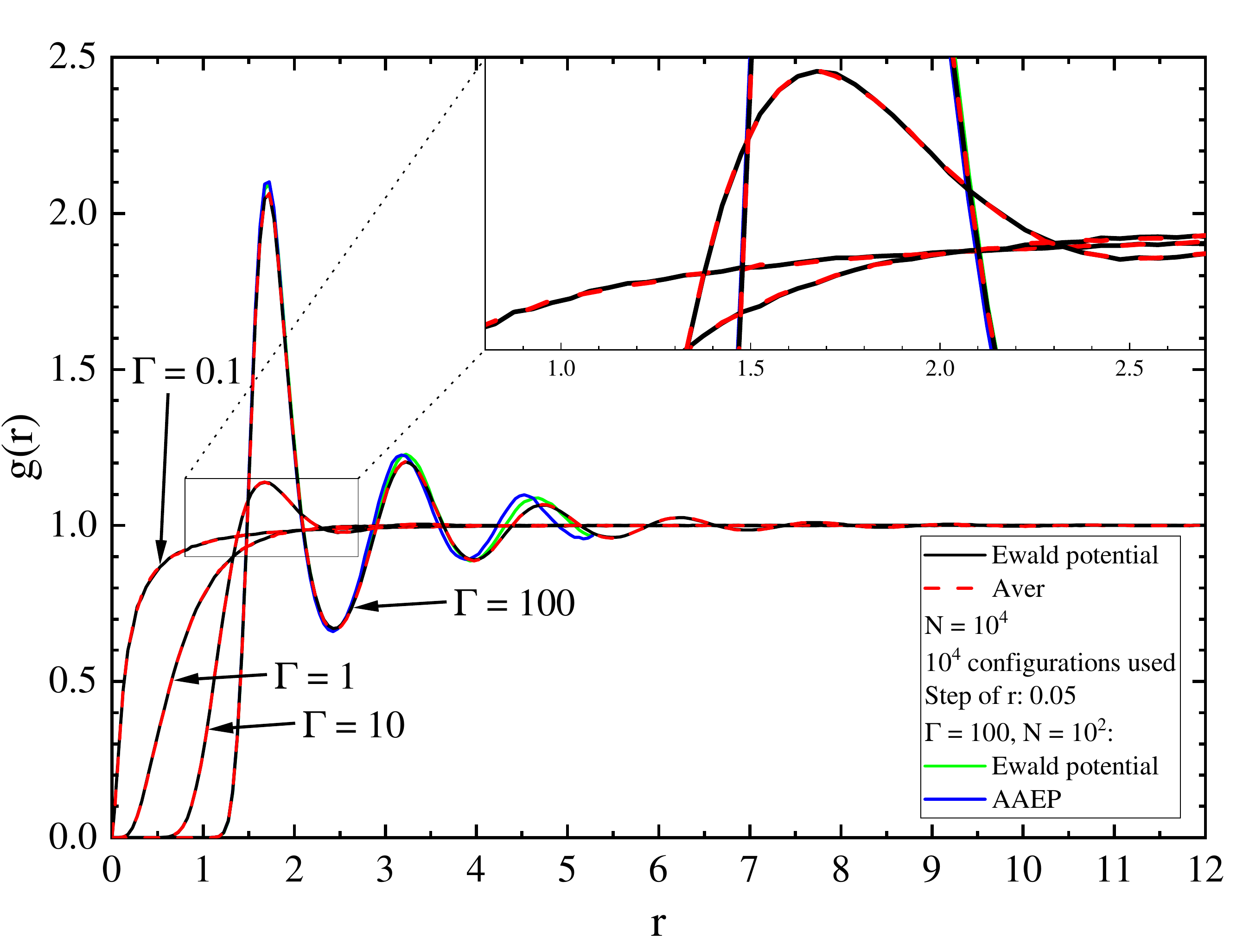}
\caption{The RDF, $g(r)$, calculated from MC simulations via the traditional Ewald potential \eqref{eq:energy} (black solid line) and the AAEP \eqref{eq:energyAv} (red dash line) for $\Gamma = 0.1, 1, 10, 100$ and $N = 10^4$. Only $10^4$ configurations are used to calculate $g(r)$. The calculations via both methods agree very well. Green and blue lines represent $g(r)$ for $\Gamma = 100, N = 10^2$ computed by both potentials; the significant difference is observed.}
\label{fig:rdf}
\end{figure}

We calculate the RDF, $g(r)$, from MC simulations via the traditional Ewald potential \eqref{eq:energy} and the AAEP \eqref{eq:energyAv} for $\Gamma = 0.1, 1, 10, 100$ and $N = 10^4$. The results are represented in Figure  \ref{fig:rdf}. For the calculation of $g(r)$ $10^4$ configurations are chosen from the MC chain. The curves obtained by both methods coincide with each other for all $\Gamma$. 

At $\Gamma = 100$ and $N = 10^2$ we observe differences in $g(r)$ starting from the second maximum (see Fig.~\ref{Fig:Ndependence}).
This behavior causes the difference in energy between the exact and angular--averaged Ewald potential. Nevertheless, both functions for $\Gamma = 100, N = 10^2$ are shifted from the correct $g(r)$. If the $N$-convergence is observed, the RDFs obtained with the Ewald and AAEP potentials coincide.

Thus, we demonstrate the applicability and high accuracy of the OCP AAEP for the calculation of energy in the broad range of the $\Gamma$ parameter. Also, the AAEP shows impressive effectiveness in comparison with the exact Ewald potential. Unfortunately, it is not possible to make  a direct performance comparison with the works \cite{Caillol:1999, Caillol:2010}.
In Ref.~\cite{Caillol:2010} it is stated that they require ``one month for 10 000 configurations'' for $N = 51200$. In our study, we generated $10^7$ configurations in 46 hours and in 3 weeks for $N = 10^5$ and $N = 10^6$, respectively. This means that a similar calculation for $N = 50000$ and $10^4$ MC steps would take $46/2/10^7\times10^4 = 0.023$ hours. Thus we estimate that the speed of our calculation is about 30 \emph{thousand} times greater than in \cite{Caillol:2010}.
Due to this fact the AAEP can be applied to a very large number of particles in a supercell. We believe that the OCP energy can be calculated by this method to refine and verify previous results for the OCP.

\section{\label{sec:conc}Conclusion}
A correct expression for the angular--averaged Ewald potential (AAEP) in an OCP was obtained. For this purpose we considered an exact anisotropic Ewald potential introduced especially for an OCP and constructed the power series of the angular--averaged potential. Using the Fourier transform, Tailor expansion and Poisson formula, we rigorously demonstrated that all coefficients of the power series are identically zero except for the first two. Then with the cluster expansion in the limit $\Gamma \to 0$ we corrected the potential to avoid the divergence of potential energy. Finally, we shifted the potential so that the resulting AAEP completely vanished at $r > r_m$. We also derived the correct expression for the OCP energy expressed through the AAEP. With the correct expression for the AAEP, it was shown that the same expression for the OCP energy can be obtained from the TCP energy. 

Our calculations of the Madelung constant demonstrated that the AAEP turned out to be about 230 times more effective than the original Ewald potential at a comparable accuracy. We also performed MC simulations in the range $0.01 \le \Gamma \le 100$ with up to a million particles in a computational cell. Our result at $\Gamma = 0.01$--$0.1$ is very close to the HNC \cite{Caillol:2010} and Ortner \cite{Ortner:1999} values. At $\Gamma = 1$ our MC data agree very well with Caillol \textit{et al.} \cite{Caillol:2010}. 

Thus, the AAEP turned out to be very effective and accurate in the simulations of classical Coulomb systems. We anticipate that the main advantage of the AAEP will show up in simulations of quantum systems such as jellium (uniform electron gas) or degenerate TCP \cite{Filinov:PRE:2015, Dornheim:PRE:2019, Dornheim:JCP:2019}.

\begin{acknowledgments}
We thank Alexander Larkin for the idea of the OCP energy derivation from the expression for the TCP energy.
\end{acknowledgments}

\appendix
\section{Coefficients of Taylor expansion}
\label{app:coefs}
Our goal now is to obtain the coefficients of the Taylor expansion \eqref{eq:taylorSeries}. A direct Taylor expansion of \eqref{eq:angularPotAveraged} can not be used, since we have the non-central terms $f(n~+~x)$ and $f(n-x)$ ($f(x)$ is defined in Eq. \eqref{eq:fDef}). This difficulty complicates the derivation compared to a TCP \cite[Eqs. (12), (13)]{Demyanov:arxiv:2022}, where these terms can be excluded.

Let us decompose $f(n+x)$  into a Fourier integral of the variable $n$:
\begin{equation}
f(n+x) = \int\limits_{-\infty}^{+\infty}\mathcal{F}_n[f(x+n)](\omega)e^{-i\omega n}d\omega,
\end{equation}
where $\mathcal{F}_n[f(n)](\omega)$ is a Fourier transform:
\begin{equation}
\mathcal{F}_n[f(n)](\omega) = \cfrac{1}{2\pi}\int\limits_{-\infty}^{+\infty}f(n)e^{i\omega n}dn.
\end{equation}
Using the following property of the Fourier transform:
\begin{equation}
\mathcal{F}_n[f(x+n)](\omega) = e^{-i\omega x}\mathcal{F}_n[f(n)](\omega),
\end{equation}
we get
\begin{equation}
f(n+x) =  \int\limits_{-\infty}^{+\infty}e^{-i\omega x}\mathcal{F}_n[f(n)](\omega).
\end{equation}
Thus, expression $f(n-x) - f(n+x)$ transforms into:
\begin{equation}
f(n-x) - f(n+x) = 2i\int\limits_{-\infty}^{+\infty}\mathcal{F}_n[f(n)](\omega)\sin(\omega x )e^{-i\omega n}d\omega.
\end{equation}
\begin{widetext}
\noindent Now we can expand $\sin(\omega x )$ into the Taylor series at $x = 0$:
\begin{equation}
\label{eq:fdhsfjd}
f(n-x) - f(n+x) = 
\sum_{k=0}^{\infty}\left(2i\int\limits_{-\infty}^{+\infty}\mathcal{F}_n[f(n)](\omega)\cfrac{(-1)^k\omega^{2k+1}}{(2k+1)!}e^{-i\omega n}d\omega\right)x^{2k+1},
\end{equation}
or, using the inverse Fourier transform notation, we rewrite Eq. \eqref{eq:fdhsfjd}:
\begin{equation}
f(n-x) - f(n+x) = 
\sum_{k=0}^{\infty}\cfrac{2i(-1)^k}{(2k+1)!}\mathcal{F}^{-1}_{\omega}[\omega^{2k+1}\mathcal{F}_n[f(n)](\omega)](n)x^{2k+1}.
\end{equation}
The Fourier transform $\mathcal{F}_n[f(n)](\omega)$ is:
\begin{equation}
\mathcal{F}_n[e^{-\pi n^2}-\pi 
n\mathrm{erfc}\left(\sqrt{\pi } n \right)] = -\frac{e^{-\pi  \omega^2}}{2 \pi  \omega^2}-\frac{1}{2} i \delta '(\omega).
\label{eq:ftf}
\end{equation}
Since we multiply \eqref{eq:ftf} by $\omega^{2k+1}$, the second term has zero impact for $k\geq 1$. Thus:
\begin{equation}
\mathcal{F}^{-1}_{\omega}\left[\omega^{2k+1}\left(-\frac{e^{-\pi  \omega^2}}{2 \pi  \omega^2}\right)\right](n) = i2^{1+2k} \pi ^{k+\frac{1}{2}} n \Gamma \left(k+\frac{1}{2}\right)
e^{-\pi n^2}M\left(1-k,\frac{3}{2},n^2 \pi \right),
\end{equation}
where $M(a,b,x)$ is the confluent hypergeometric function defined by the series:
\begin{equation}
\label{eq:hypergeometricDef}
M(a, b, x) = 
\sum _{{s=1}}^{\infty }{\frac{a^{{(s)}}}{b^{{(s)}}s!}}x^{s},
\end{equation}
where $a^{{(s)}}$ denotes the rising factorial:
\begin{equation}
a^{(0)}=1, a^{{(s)}}=a(a+1)(a+2)\cdots (a+s - 1).
\end{equation}
Then the second part of the pair the AAEP $v^a_2(x)$ has the following form:
\begin{equation}
Lv^a_2(x) = \sum_{k=0}^{\infty}(-1)^k\sum_{\textbf{n}\neq\textbf{0}}e^{-\pi n^2}\left[
-\cfrac{2^{1+2k} \pi ^{k-\frac{1}{2}}}{(2k+1)!}
\Gamma \left(k+\frac{1}{2}\right)
M\left(1-k,\frac{3}{2},n^2 \pi \right)
+ \cfrac{2^{2 k} \pi^{2 k-1}  n^{2 k-2}}{(2 k+1)!   }
\right]x^{2k}.
\end{equation}
Expanding the $\mathrm{erfc}$-function at $x = 0$ in Eq. \eqref{eq:spherEwald}, we get the expression for $C_k$ in Eq.~\eqref{eq:taylorSeries} (if $k\geq 1$):
\begin{equation}
C_k = (-1)^k\sum_{\textbf{n}\neq\textbf{0}}e^{-\pi n^2}\left[
 \cfrac{2^{2 k} \pi^{2 k-1}  n^{2 k-2}}{(2 k+1)!   }
-\cfrac{2^{1+2k} \pi ^{k-\frac{1}{2}}}{(2k+1)!}
\Gamma \left(k+\frac{1}{2}\right)M\left(1-k,\frac{3}{2},n^2 \pi \right)
\right] 
-\frac{2 (-1)^k \pi ^k}{(2 k+1) k!}.
\end{equation}
By adding and subtracting the contribution at $\textbf{n} = \textbf{0}$ and using the $\Gamma$-function property $\Gamma(k+1/2) = (2n)!\sqrt{\pi}/(2^{2n}n!)$, we get the final expression for $C_k$:
\begin{equation}
\label{eq:CoefsExpression}
C_k = (-1)^k\sum_{\textbf{n}}e^{-\pi n^2}\left[
\cfrac{2^{2 k} \pi^{2 k-1}  n^{2 k-2}}{(2 k+1)!   }
-\cfrac{2^{1+2k} \pi ^{k-\frac{1}{2}}}{(2k+1)!}
\Gamma \left(k+\frac{1}{2}\right)
M\left(1-k,\frac{3}{2},n^2 \pi \right)
\right] +
\cfrac{2 \pi }{3  }\delta_{1,k}, k\geq 1,
\end{equation}
where $\delta_{1,k}$ is the Kronecker symbol.
\section{Zeroing the coefficients}
\label{app:summation}
We are going to prove the equality:
\begin{equation}
\label{eq:proof1}
\cfrac{2^{2 k} \pi^{2 k-1}  }{(2 k+1)!   }\sum_{\textbf{n}}e^{-\pi n^2}n^{2 k-2}
=
\cfrac{2^{1+2k} \pi ^{k-\frac{1}{2}}}{(2k+1)!}
\Gamma \left(k+\frac{1}{2}\right)\sum_{\textbf{n}}e^{-\pi n^2}
M\left(1-k,\frac{3}{2},n^2 \pi \right),
\end{equation}
from which it follows that $C_k = 0$ for $k\geq 2$. To do this, we use the Poisson formula: 
\begin{equation}
\sum_{\textbf{n}}e^{-\pi n^2}n^{2 k-2} = \sum_{\textbf{q}}F_k(\textbf{q}),
\end{equation} 
where
\begin{equation}
F_k(\textbf{q}) = \int e^{-\pi n^2}n^{2 k-2} e^{-2\pi i \textbf{n}\cdot\textbf{q}}d^3n = 2\pi^{\frac{1}{2}-k}\Gamma\left(k+\cfrac{1}{2}\right)e^{-\pi q^2}M\left(1-k,\frac{3}{2},q^2 \pi \right)
\label{eq:poisson}
\end{equation}
is a Fourier transform. Substituting \eqref{eq:poisson} in the left--hand side of Eq.~\eqref{eq:proof1} turns Eq.~\eqref{eq:proof1} into an identity. Thus, $C_k = 0$ for $k\geq 2$.
\end{widetext}
\section{Derivation the OCP AAEP from the TCP one}
\label{app:derivocptcp}
The main idea of this derivation was suggested by Alexander Larkin.

In Refs.~\cite{Yakub:2003}, \cite[Eq. (59)]{Demyanov:arxiv:2022} the energy of TCP was obtained by the angular averaging of the TCP Ewald potential. We consider a plasma with two sorts of particles with charges $Q_+~=~+Ze$ and $Q_-~=~-Ze$. Then the internal energy reads:
\begin{equation}
\label{eq:energytcp}
\cfrac{U}{\Gamma} = -\sum_{i = 1}^N\cfrac{3z_i^2}{4r_m} + \cfrac{1}{2}\sum_{i = 1}^Nz_i\sum_{\substack{j=1\\ j\neq i}}^{N} z_j\tilde{v}(r_{ij}).
\end{equation}
Here, $\Gamma = (Ze)^2/(k_BT a)$, $z_i$ is a charge \emph{sign} and $\tilde{v}(r)$ is defined by Eq. \eqref{eq:shiftedPot}. Below we obtain Eq. \eqref{eq:energyAv} for the OCP energy from Eq. \eqref{eq:energytcp}. The main idea is to separate the summation in Eq.~\eqref{eq:energytcp} for positive and negative ions; then consider a transition from the summation over point charges to the integration over a uniformly distributed charge.

We separate positive and negative ions:
\begin{equation}
N = N_+ + N_- = 2N_+ = 2N_-,
\end{equation}
where $N_+$ and $N_-$ denote the number of positive and negative ions, respectively. Let us consider the first term in Eq. \eqref{eq:energytcp}:
\begin{equation}
\label{eq:appqweqw}
-\sum_{i = 1}^N\cfrac{3z_i^2}{4r_m} = -\sum_{i = 1}^{N_-}\cfrac{3z_i^2}{4r_m}-\sum_{i = 1}^{N_+}\cfrac{3z_i^2}{4r_m}.
\end{equation}
For positive ions, we keep this term as a discrete sum with $z_i = 1$:
\begin{equation}
\label{eq:app:1}
-\sum_{i = 1}^{N_+}\cfrac{3z_i^2}{4r_m} = -\cfrac{3N_+}{4r_m}.
\end{equation}
Now we consider each negative ion as a space element with an infinitesimal charge
\begin{equation}
z_i \rightarrow -\rho dV,
\end{equation}
where $\rho = N_+/L^3 = N_-/L^3$ is the \emph{number} density and $dV$ denotes a space element. Then we replace the summation over charges to an integral over space:
\begin{equation}
\sum_{i = 1}^{N_-}z_i \rightarrow -\int\rho dV.
\end{equation}

The last term in Eq. \eqref{eq:appqweqw} contains $z^2_i$, while the summation is performed once. As we proceed to integration, this term has an infinitesimal contribution to the energy of the system:
\begin{equation}
\label{eq:app:2}
-\sum_{i = 1}^{N_-}\cfrac{3z_i^2}{4r_m} \rightarrow 0.
\end{equation}

Next, we consider the two-particle contribution to the energy. This term results in three sums:
\begin{multline}
\label{eq:fdfgfdg}
\sum_{i = 1}^Nz_i\sum_{\substack{j=1\\ j\neq i}}^{N} z_j\tilde{v}(r_{ij}) = 
\sum_{i = 1}^{N_-}z_i\sum_{\substack{j=1\\ j\neq i}}^{N_-} z_j\tilde{v}(r_{ij})
\\+
2\sum_{i = 1}^{N_+}z_i\sum_{j = 1}^{N_-} z_j\tilde{v}(r_{ij})
+
\sum_{i = 1}^{N_+}z_i\sum_{\substack{j=1\\ j\neq i}}^{N_+} z_j\tilde{v}(r_{ij}).
\end{multline}
The last term in Eq. \eqref{eq:fdfgfdg} remains unchanged. We change the second term into the integral over space: 
\begin{multline}
2\sum_{i = 1}^{N_+}z_i\sum_{j = 1}^{N_-} z_j\tilde{v}(r_{ij}) \rightarrow 
-2\sum_{i = 1}^{N_+}z_i\int \rho\tilde{v}(|\textbf{r}_j - \textbf{r}_i|)d\textbf{r}_j
\\=
-2\sum_{i = 1}^{N_+}z_i\int \rho\tilde{v}(|\textbf{r}_j - \textbf{r}_i|)d(\textbf{r}_j - \textbf{r}_i)
 \\=
-2\sum_{i = 1}^{N_+}z_i\int\limits_{0}^{r_m} \rho\tilde{v}(u)d\textbf{u},
\end{multline}
where $\textbf{u} = \textbf{r}_j - \textbf{r}_i$.
We calculate this integral in spherical coordinates:
\begin{multline}
-2\sum_{i = 1}^{N_+}z_i\int\limits_{0}^{r_m} \rho\tilde{v}(u)d\textbf{u} = 
-2\sum_{i = 1}^{N_+}z_i\int\limits_{0}^{r_m} 4\pi u^2\rho\tilde{v}(u)du 
\\= 
-2N_+ 4\pi\rho r^2_m \cfrac{1}{10}.
\end{multline}
Since
\begin{equation}
\label{eq:densityapp}
\rho = \cfrac{N_+}{L^3} = \cfrac{N_+}{\tfrac{4\pi}{3}r_m^3},
\end{equation}
we obtain the expression:
\begin{equation}
\label{eq:app:3}
2\sum_{i = 1}^{N_+}z_i\sum_{j = 1}^{N_-} z_j\tilde{v}(r_{ij}) \rightarrow 
-\cfrac{6}{10}\cfrac{N_+^2}{r_m}.
\end{equation}

The first term in Eq. \eqref{eq:fdfgfdg} is also replaced with an integral:
\begin{equation}
\sum_{i = 1}^{N_-}z_i\sum_{\substack{j=1\\ j\neq i}}^{N_-} z_j\tilde{v}(r_{ij}) \rightarrow
-\sum_{i = 1}^{N_-}z_i 4\pi \rho \int\limits_{0}^{r_m}u^2\tilde{v}(u)du.
\end{equation}
We calculate this integral and get:
\begin{equation}
-\sum_{i = 1}^{N_-}z_i 4\pi \rho r_m^2/10
\end{equation}
This sum also should be integrated. Since the integrand is constant, we obtain:
\begin{equation}
-\sum_{i = 1}^{N_-}z_i 4\pi \rho r_m^2/10
\rightarrow
\rho \times\cfrac{4\pi}{3}r^3_m \times 4\pi \rho r_m^2/10
\end{equation}
Using Eq. \eqref{eq:densityapp}, we get:
\begin{equation}
\label{eq:app:4}
\sum_{i = 1}^{N_-}z_i\sum_{j = 1\atop j\neq i}^{N_-} z_j\phi(r_{ij}) \rightarrow
\cfrac{3}{10}\cfrac{N_-^2}{r_m} = \cfrac{3}{10}\cfrac{N_+^2}{r_m}.
\end{equation}

Substituting Eqs.~\eqref{eq:app:1}, \eqref{eq:app:2}, \eqref{eq:fdfgfdg}, \eqref{eq:app:3}, \eqref{eq:app:4} in \eqref{eq:energytcp}, we get the OCP energy:
\begin{multline}
\cfrac{U}{\Gamma} = -\cfrac{3N_+}{4r_m}-\cfrac{1}{2}\cfrac{6}{10}\cfrac{N_+^2}{r_m}+\cfrac{1}{2}\cfrac{3}{10}\cfrac{N_+^2}{r_m} + \cfrac{1}{2}\sum_{i = 1}^{N_+}\sum_{\substack{j=1\\ j\neq i}}^{N_+}\tilde{v}(r_{ij})
\\=
-\cfrac{3}{20}\cfrac{N_+(N_++5)}{r_m} + \cfrac{1}{2}\sum_{i = 1}^{N_+}\sum_{\substack{j=1\\ j\neq i}}^{N_+} \tilde{v}(r_{ij}).
\end{multline}
This expression is the same as Eq.~\eqref{eq:energyAv} obtained from the OCP AAEP.

\providecommand{\noopsort}[1]{}\providecommand{\singleletter}[1]{#1}%
%

%\bibliography{refs}
\end{document}